\documentclass{article}

\usepackage{arxiv}

\usepackage[utf8]{inputenc} 
\usepackage[T1]{fontenc}    
\usepackage{hyperref}       
\usepackage{url}            
\usepackage{booktabs}       
\usepackage{nicefrac}       
\usepackage{microtype}      
\usepackage{lipsum}		
\usepackage{graphicx}
\usepackage[super,comma,sort&compress]{natbib} 
\usepackage{doi}
\usepackage{amsmath}
\usepackage{bbm}
\usepackage{bm}
\usepackage{color}
\usepackage{siunitx}
\usepackage{subcaption} 
\usepackage{booktabs} 
\usepackage{multirow} 
\newcommand{\E}{\text{E}}
\newcommand{\NA}{\text{-}}
\newcommand{\Var}{\text{Var}}
\newcommand{\Corr}{\text{Corr}}
\newcommand{\Cov}{\text{Cov}}
\newcommand{\Prob}{\text{P}}
\newcommand{\pkg}{\textbf}

\title{A comparison between copula-based, mixed model, and estimating equation methods for regression of bivariate correlated data}

\author{ \href{https://orcid.org/0009-0002-6621-8793}{\includegraphics[scale=0.06]{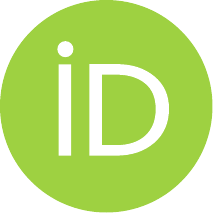}\hspace{1mm}Aydin Sareff-Hibbert} \\
	Faculty of Medicine and Health\\
	University of Sydney\\
	NSW, Australia \\
	\texttt{aydin.hibbert@sydney.edu.au} \\
	\And
	\href{https://orcid.org/0000-0003-1270-1499}{\includegraphics[scale=0.06]{orcid.pdf}\hspace{1mm}Gillian Z. Heller} \\
	Faculty of Medicine and Health\\
	University of Sydney\\
	NSW, Australia \\
	\texttt{gillian.heller@sydney.edu.au} \\
}

\renewcommand{\shorttitle}{A comparison of methods for regression of bivariate correlated data}

\hypersetup{
pdftitle={A comparison between copula-based, mixed model, and estimating equation methods for analysis of bivariate correlated data},
pdfsubject={q-bio.NC, q-bio.QM},
pdfauthor={Aydin Sareff-Hibbert, Gillian Z. Heller},
pdfkeywords={generalized joint regression, random effects, mixed models, copula, correlated data},
}

\begin{document}
\maketitle

\begin{abstract}
This paper presents a simulation study comparing the performance of generalized joint regression models (GJRM) with generalized linear mixed models (GLMM) and generalized estimating equations (GEE) for regression of longitudinal data with two measurements per observational unit. We compare models on the basis of overall fit, coefficient accuracy and computational complexity. 
\\ \\
We find that for the normal model with identity link, all models provide accurate estimates of regression coefficients with comparable fit. However, for non-normal marginal distributions and when a non-identity link function is used, we highlight a major pitfall in the use of GLMMs: without significant adjustment they provide highly biased estimates of marginal coefficients and often provide extreme fits. GLMM coefficient bias and relative lack of fit is more pronounced when the marginal distributions are more skewed or highly correlated. In addition, we find major discrepancies between the estimates from different GLMM software implementations. In contrast, we find that GJRM provides unbiased estimates across all distributions with accurate standard errors when the copula is correctly specified; and the GJRM provides a model fit favourable or comparable to GLMMs and GEEs in almost all cases. We also compare the approaches for a real-world longitudinal study of doctor visits. 
\\ \\ 
We conclude that for non-normal bivariate data, the GJRM provides a superior model with more consistently accurate and interpretable coefficients than the GLMM, and better or comparable fit than both the GLMM and GEE, while providing more flexibility in choice of marginal distributions, and control over correlation structure.
\end{abstract}


\section{Introduction}

Dependence between random variables often develops when repeated observations are taken from a sampling unit, for example, cholesterol levels over time for a single patient. If independence is assumed between these correlated observations, the observations will exhibit a lower than expected variance, which causes bias and inefficiency in coefficient estimation \citep{Laird1982}.

Several approaches have been developed to date to account for the dependence between observations in regression, the most popular in the case of non-normal data being generalized linear mixed models (GLMM)\citep{Breslow1993} and generalized estimating equations (GEE)\citep{Liang1986}. These methods introduce an adjustment to univariate methods that account for the dependence structure between observations. Both frameworks have been substantially extended since their inception to apply to a much broader range of response distributions, as well as to incorporate smooth terms for covariates\citep{hastie1990generalized,Stasinopoulos2017,Yee1996,Wild1996}. 

Both the GEE and GLMM are based on the framework of the Generalized Linear Model (GLM)\citep{McCullaghNelder89} and introduce dependence to the GLM via different methods. For a given longitudinal dataset $y_{it}$ observed at times $t=1,\ldots,T$ and for subjects $i=1,\ldots,n$, the standard GLM assumes within-subject (and between-subject) independence. For the GLM, we have the model: 
$$g(\mu_{it})=\eta_{it}=x_{it}^T\beta$$
where $g(\cdot)$ is the link function that connects the linear predictor $\eta_{it}$ to the response parameter $\mu_{it}$ and $\beta$ is the parameter vector that links the covariates $x_{it}^T$ to the linear predictor.

The GEE uses the same model base as the GLM but introduces a working correlation matrix among measurements for a subject (or cluster) $i$, $R_i$, with assumed independence between subjects. This correlation structure is used to provide adjusted population and individual-level standard errors of parameter estimates to account for correlation. Given the flexibility of the covariance structure, multiple structures can be modeled, popularly: a single correlation parameter $\rho$ for all observations in a cluster, referred to as an exchangeable correlation structure; AR1 with a single correlation parameter but with correlation decaying with increasing distance; or fully unstructured. Modern software implementations of the GEE\citep{vanegas_generalized_2023} extend the available correlation structures further and provide additional tools for model selection, diagnostics and more complex datasets or covariates. This method provides accurate estimates for marginal coefficients while providing more accurate standard errors than the comparable GLM for a correlated dataset. Importantly, one significant drawback of the GEE is that it does not utilize a valid likelihood, so resulting estimates are not maximum likelihood, and standard likelihood-based model selection criteria such as Akaike Information Criterion (AIC) are not available.

The GLMM introduces dependence between subjects by introducing random intercept and/or random slope terms into the linear predictor, which provide a common value for all observations of a given subject. The simplest and most commonly used form of the GLMM is that of a single random intercept term which induces dependence within subjects by introducing a normally distributed random effect term centred at zero, $b_i\sim \mathcal{N}(0,\theta^2),\quad i=1,\ldots,n$, into the linear predictor of the GLM, which is fixed across all observations for subject $i$:
$$g(\mu_{it})=\eta_{it}=x_{it}^T\beta+b_i.$$
The introduction of the random effect term changes the structure of the data being modeled as instead of the distributional model directly modeling the marginal response distributions as in the case of the GLM and GEE, i.e. $Y_{t} \sim \mathcal{D}(\mu_t,\sigma_t)$, the distribution being modeled is conditional on the random effect, $Y_t | b \sim \mathcal{D}(\mu_t,\sigma_t)$.

In the case of a mixed model with a random effect term, the way parameters are interpreted needs to be carefully considered due to the impact of the random effect on the parameter estimates of the model \citep{gory_class_2021}. This is because there is a difference between the marginal and conditional estimates in the GLMM framework. For example, if the means of the marginal distributions are the values of interest, i.e. $E(Y_t)$, then when $g(\cdot)$ is the identity link function, the conditional and unconditional expectations for the means are equivalent because the expected value of the random effect is zero and can be dropped, i.e.
\begin{align*}
E_b[E_Y(Y_t|b_i)]=E_b\left[ x_{it}^T\beta+b_i \right] = x_{it}^T\beta + E_b(b_i)=E(Y_t).
\end{align*}

However, in the case where the link function is not identity, then $E_b[E_Y(Y_t|b_i)]\neq E_Y(Y_t)$ as the link function prevents the random effect from being dropped and needs to be evaluated over the random effect, i.e. 
\begin{align*}
E_b[E_Y(Y_t|b_i)]=E_b\left[ g^{-1}\left(x_{it}^T\beta+b_i\right) \right] \neq E(Y_t),
\end{align*}

in general.

This means both the shape of the distribution being modeled and the parameter estimates from the GLMM are distorted by the random effect, and as the parameters are based on the conditional expectation this makes interpretation problematic. This issue of parameter interpretation exists across all popular non-identity link functions\citep{gory_class_2021}. 

Many solutions have been proposed to the problem of providing better marginal interpretability for GLMMs. These solutions include: specifying the models for marginal interpretation, termed marginalized multi-level models, by incorporating an additional adjustment term in the linear predictor for each observation\citep{gory_class_2021,heagerty_marginalized_2000,mills_marginally_2002}; the use of bridging distributions such that the integral of the predictor $h(\eta+b)$ with respect to the bridging distribution provides marginal parameters \citep{wang_matching_2003}; and selecting a random effect distribution which is conjugate to the outcome distribution which can provide closed form solutions to estimates for marginal joint distributions and marginal parameters for many exponential family distributions\citep{molenberghs_family_2010}. Helpful overviews of the relationship between marginal, bridging and conjugate solutions are provided by Kenward and Molenberghs\citep{kenward_taxonomy_2016} and Molenberghs et al.\citep{molenberghs_connections_2013}. These methods generally introduce additional restrictions to analysis, for example restricting to only exponential family distributional forms, and complexity to the construction and estimation of the model, largely due to the need for integration over often complex random effect components and the adjustments to the model structure to account for them. 

In practice, these adjustment methods are rarely applied; rather, the simplifying assumption that $\E(Y_t|b_i)=\E(Y_t)$ is used, generally without clarification, and inference is based on the results from standard packages which provide conditional results. For this reason, in this paper we analyze the results of a standard GLMM implemented by common packages without the major adjustments set out above. To highlight this point of how commonly GLMMs are used without adjustment, we performed a non-comprehensive review of recent papers utilizing GLMMs in the life sciences. We use the search term "generalized linear mixed models" in PubMed and review the 20 most recent papers of 252 published in 2024, resulting in papers with create dates from October 26 to November 21, 2024. Our interest is in the total volume and breadth of the articles, which is significant, the complexity of the structure used, and whether complex adjustments are made to account for the difference between marginal and conditional results.

Of the twenty papers reviewed, eight were non-normal GLMMs with a single random intercept. Of the others, two had two or more random effects, two were linear models (normal response with identity link), two were meta-analyses, three were not studies including GLMMs, and three had more complex structures. None applied the complex adjustments outlined above. The papers model a broad range of measures in health and life sciences including: association between 24-hour movement guidelines and academic engagement in adolescents \cite{sanchez-miguel_unveiling_2024}, sex differences in survival following acute coronary syndrome \cite{anand_sex_2024}, fetal risk for congenital heart defects \cite{pi_periconceptional_2024}, increasing HIV testing uptake using mobile technology \cite{schnall_efficacy_2024}, novel soft contact lenses impact on patient comfort \cite{buch_comfort_2024}, antenatal care utilization in Zambia\cite{shumba_spatial-temporal_2024}, negative affect and loss of control eating following bariatric surgery\cite{kerver_naturalistic_2025}, and the association between residential greenness and obesity \cite{yu_associations_2024}. More broadly, there are applications across life sciences including in nutrition \cite{mullertz_cross-sectional_2024}, medicine \cite{rysavy_assessment_2020},\cite{chopra_michigan_2017}, and fisheries \cite{gundelund_investigating_2022}.

Many authors highlight the rapidly increasing use of GLMMs to analyze correlated data in research papers in the life sciences over the past 20 years and provide reviews of their use specifically in clinical medicine \citep{casals_methodological_2014}, psychology \citep{bono_report_2021}, and biology \citep{bolker_generalized_2009,pekar_generalized_2018,thiele_potential_2012}. Interestingly, while the GEE provides a similar adjustment for correlation, it has not seen anywhere near the increase in adoption as GLMMs \citep{pekar_generalized_2018}. 
In the vast majority of cases, the GLMM is applied simply for the purpose of adjusting for known correlation between observations within clusters, that is, to adjust fixed effect parameters, essentially treating the correlation as a nuisance parameter \citep{thiele_potential_2012, pekar_generalized_2018}. 
Most of the studies in these reviews use a single random intercept\citep{casals_methodological_2014,pekar_generalized_2018}, which is why we focus on this structure throughout this paper. The most popular model error structures used in the reviewed papers were Poisson, Binomial, and Bernoulli. 

The increased complexity of the GLMM structure introduces additional risks to the modeling process that are difficult to account for even for an experienced analyst \citep{silk_perils_2020,bolker_generalized_2009,arnqvist_mixed_2020}. For example, one analysis\cite{bolker_generalized_2009} finds 58 percent of reviewed papers used GLMM inappropriately in some way, and others\cite{casals_methodological_2014,bono_report_2021} note that the majority of articles reviewed did not report critical model information needed to assess model validity, or even the distribution and link function in many cases.

\textbf{Copula regression}

More recently, approaches for joint modeling of multivariate dependent data in a regression framework have emerged through the use of copulas, which provide an alternative to the former models. In particular, generalized joint regression models (GJRM)\citep{Marra2017} provide a likelihood-based framework to jointly model multiple correlated variables and their dependence structures while incorporating covariate effects. The GJRM combines the additional utility of the likelihood-based approach of the GLMM with the marginal interpretability of the GEE.

The GJRM framework relies on the use of copulas. Copulas provide a convenient method for deconstructing a bivariate distribution into a combination of two marginal distributions and a copula function to model the dependence structure \citep{Nelsen2007,Trivedi2007}. Any continuous bivariate distribution can be represented by a unique combination of two marginal distribution functions and a copula function defining the dependence structure, and this extends readily to the multivariate case \citep{sklar1973random}. Discrete distributions can also be deconstructed in the same way with copulas; however, the resulting distribution is not unique. Copulas have been widely used in economics and financial time series analysis, particularly in portfolio risk \citep{ParraPalaro2006, Pitt2006} and insurance loss estimation \citep{Kramer2013}. 

One of the key developments in copula regression is the introduction of a framework for conditioning a copula on a variable (covariate)\citep{Patton2006}. This allows dependence structures to be modeled after the effects of covariates are removed, leaving the true residual dependence structure and significantly expanding the applications of copulas in a regression context. This method is referred to as conditional copula regression. A large amount of literature has focused on the development of likelihood-based tests for the existence and significance of covariates in a conditional copula regression framework \citep{Acar2011, Gijbels2011, Acar2013,Craiu2012,Sabeti2014}.

The first approaches to fully flexible joint regression with regression covariates for both the margins and dependence are limited to modelling the bivariate case, and for some specific distributions, trivariate. These approaches are the Generalized Joint Regression Model (GJRM)\citep{marra_copula_2025} and the Generalized Additive Model for conditional dependence structures (gamCopula) \citep{Vatter2015}. Both approaches combine univariate models for marginal distributions, generalized additive models for location, scale and shape, GAMLSS\citep{stasinopoulos_generalized_2024}, and generalized additive models, GAM\citep{hastie1990generalized}, respectively, with a fit for a copula to capture the dependence structure. The methods differ in their approach to optimization of the joint likelihood function: the \textbf{R}\citep{R2024}  package \pkg{GJRM} \cite{Marra2017} uses a simultaneous estimation approach for the joint likelihood of the copula and marginal distributions, and \pkg{gamCopula} \citep{Vatter2015} maximizes the marginal and copula likelihoods separately. The GJRM supports many, but not all, of the broad range of marginal distributions provided by GAMLSS. In our analysis we will mainly focus on the use of GJRM over gamCopula as the implementations provide similar results for the copula parameters but the GJRM delivers slightly less biased and more efficient estimates for marginal parameters than the approach using gamCopula, and has faster computation\citep{Marra2017}. In addition, the GJRM implementation has been substantially extended since its development, with detailed support having been published\citep{marra_copula_2025}. A Bayesian approach for simultaneous estimation in a bivariate copula regression has also been introduced; this provides another alternative to the former methods\cite{Klein2016}, however we focus this analysis on likelihood- and quasi-likelihood methods due to their relative popularity and computational advantages. Future reviews may focus on further comparisons. Another popular package for copula-based regression is \pkg{VineCopula}\citep{czado_vine_2022} which provides a method for modelling multivariate longitudinal data with arbitrary margins by combining bivariate copulas to describe multivariate dependence. However this approach does not take into account covariate fits for the dependence structure, and does not jointly fit copula and margin as is the case for the GJRM.

Using a copula-based joint regression model such as GJRM, we can specify a longitudinal model comparable to that of the GEE or GLMM, with a distributional fit for each margin and a likelihood-based dependence structure fit through the copula.

Adjustments for dependence incorporated in GLMMs and GEEs are essentially approximate approaches for accounting for multivariate data features in a univariate regression framework. In contrast, a copula-based joint regression approach directly models the multivariate distribution, with fits for observed margins and dependence structure, allowing for a high level of control and transparency in describing the underlying data. The structure also allows for a high level of flexibility in use of the final model, for example, being able to analyze the fitted tails of the margin or copula distribution.

\subsection{Contribution of this paper}

To date, applications of joint regression models that use covariate-dependent copulas and margins have been mainly limited to the bivariate case and have focused mainly on modeling different outcome variables simultaneously, for example, simultaneous modeling of poverty and leisure time \citep{Marra2023} or health expenditure and health outcomes \citep{Dorn2023}. We review an alternative potential use for copula-based joint regression models for the case of longitudinal data regression, where the same outcome variable is measured at multiple time points for the same cluster or subject. We are not aware of any previous research which has used copula-based joint regression models for regression of longitudinal data and compared them to existing methods.

This paper focuses on contrasting the performance of copula-based joint regression models, in particular the generalized joint regression model (GJRM) \cite{marra_copula_2025}, with the most popular methods for adjusting for dependence in regression: the incorporation of a single random intercept term in standard generalized linear mixed models (GLMM), and generalized estimating equations (GEE). 

Where possible, we simulate bivariate distributions which originate from neither random effects nor a copula, so as not to bias the estimation towards either of these models. We compare model performance on the basis of: overall fit through likelihood-based model selection and weighted variogram scores, accuracy of coefficient estimates and standard error, and computational complexity evaluated on runtime and convergence.

We limit our investigation to the simplest case of the estimation of coefficients for two correlated outcome variables having the same distributional form dependent on covariates. This corresponds to a longitudinal model with two time points, so there are two measurements per observational unit. The restriction to bivariate data is largely due to the limitation of the \pkg{GJRM} software, which provides methods for only bivariate and, for a limited set of distributions, trivariate joint regression. 

We demonstrate the following key points throughout the following sections:
\begin{itemize}
    \item For the normal distribution with identity link, all models (GJRM, GLMM, GEE and GLM) provide accurate coefficient estimates and standard errors.
    \item For non-normal distributions, with non-identity link, the use of random intercepts in GLMMs to adjust for correlation, without major adjustment to model structure, provides highly biased estimates of the marginal intercept coefficients, and covariate coefficients to a lesser extent depending on the distribution. This bias increases with greater marginal skewness and rank correlation. We also show substantially differing estimates of standard error and stability between standard GLMM implementations, including the \textbf{R} packages \pkg{lme4}\citep{lme4pack}, \pkg{gamlss}\citep{gamlsspack}, and \pkg{mgcv}\citep{mgcv}. In contrast, we show that the GJRM and GEE methods provide accurate and stable coefficient estimates and standard errors. 
    \item We highlight that the GJRM provides overall model fit comparable to the GLMM by AIC and Generalized AIC (GAIC), and superior by Bayesian Information Criterion (BIC) to any other compared model. In addition, we show that the GJRM provides better fits on the basis of weighted variogram scores than all other models when a non-identity link function is required. In general, the GEE also provides good fits on the basis of weighted variogram scores. 
    \item We demonstrate in a real world dataset of doctor visits over time which are both zero-inflated and skewed, that the greater range of marginal distributions available to GJRM, which incorporates many, but not all, of the GAMLSS distributions, allows it to provide a substantially better model fit than the GEE and GLMM (\pkg{lme4} and \pkg{mgcv}); and a more easily interpretable model than the standard GLMM even with the same marginal distribution implemented by \pkg{gamlss}.
\end{itemize}

The key finding we highlight for the GLMM is a pitfall of simply applying a random intercept term to adjust for correlation without carefully accounting for the change in interpretation it implies. We suggest that the copula-based approach provides a more directly controllable and transparent approach to capturing correlation in longitudinal datasets. The fact that it retains likelihood-based tools such as AIC is an advantage over the GEE.

There is a potential for copula-based joint regression approaches for longitudinal data to be extended to the multivariate case, with a mathematical framework for joint regression using copulas in more than two dimensions having been developed\cite{Kock2023} but not available in standard packages. However, analysis of the performance of copula-based models in more than two dimensions falls outside the scope of this investigation.

\section{Simulation}
This paper adopts a simulation approach to compare the performance of flexible copula-based joint regression with alternative methods to model correlated data. We simulate the simplest case of a longitudinal dataset with two time points.

Four bivariate distributions are simulated to capture different distributional qualities: a bivariate normal for continuous unbounded margins and a symmetric dependence structure, a bivariate negative binomial for discrete positive margins with mixed skewed and symmetric dependence, a bivariate Gamma for continuous positive margins with a skewed to highly skewed dependence structure, and a bivariate Bernoulli to provide a binary marginal distribution. For the non-normal distributions simulated, we simulate bivariate distributions which originate from neither random effects nor a copula, so as not to bias the estimation towards either model. 

Simulations of each distribution are run across a range of parameters of the distributions to capture as many varied data shapes as possible, resulting in 1,150 total simulated bivariate distributional shapes across the four distributions. The models include two covariates: $x_1$, which is  binary  with $\beta_{x_1}=1$, and a continuous covariate, $x_2\sim \text{Uniform}(1, 100)$,  with $\beta_{x_2}=0.01$.

Across the simulations, we use Kendall's $\tau$ to describe the strength of dependence between the random variables for the response at both time points, $Y_1$ and $Y_2$, and marginal skewness to provide additional information on distribution shape, calculated as the skewness for $Y_1$ and $Y_2$ divided by two. 

Kendall's $\tau$ represents the degree of concordance between two random variables with a value ranging between -1 and 1 and is calculated as
$$\tau=2\,\text{P}[(Y_1-Y_1')(Y_2-Y_2')>0]-1 ,$$
where $(Y_1',Y_2')^\top$ is an independent realization of the random variables describing the random vector $(Y_1,Y_2)^\top$.
As a measure of dependence, Kendall's $\tau$ has the advantage that it is not dependent on the relative scale of observations for continuous distributions, as is the case with Pearson correlation. 

The definition of the bivariate distributions used is included in the following section. 

\subsection{Bivariate distributions} \label{sec:bivariate_distributions}

\subsubsection{Bivariate normal}

Much of the theory of random effect models and generalized estimating equations was developed around the multivariate normal distribution, which we use as a benchmark for the performance of the models. We specify the five-parameter bivariate normal model as: 
\begin{align*}
\begin{pmatrix}    
Y_1\\Y_2
\end{pmatrix}\sim\mathcal{N}_2\left(\begin{pmatrix}    
\mu_1\\\mu_2
\end{pmatrix},
\begin{pmatrix}
\sigma_1^2&\rho\sigma_1\sigma_2\\  
\rho\sigma_1\sigma_2&\sigma_2^2
\end{pmatrix}
\right)
\end{align*}
having $\E(Y_t) = \mu_t$, $\Var(Y_t) = \sigma_t^2$ and $\Corr(Y_1,Y_2) = \rho$, for $t=1, 2$.

We generate 225 realizations of the distribution with 1,000 observations at both time points with fixed marginal means of $\mu_1=1$ and $\mu_2=2$; varying values of $\sigma_1$ and $\sigma_2$ between 0.25 and 2.5; and correlation between 0.1 and 0.9. The \textbf{R} package \pkg{mvtnorm}\cite{mvtnorm} is used for the random number generation.

\subsubsection{Bivariate negative binomial}

We use a bivariate compound Poisson model\cite{stein_bivariate_1987} to introduce a skew to our bivariate simulations, as well as a non-Gaussian dependence structure. This is constructed as

$$Y_1|\lambda, t_1\sim P(\lambda t_1)\quad\text{independently of }\quad Y_2|\lambda, t_2\sim P(\lambda t_2)$$
and
$$\lambda\sim \text{Gamma}(\mu,\sqrt{\sigma}),$$
where the above Gamma distribution uses the GAMLSS parameterization \texttt{GA}\citep{rigby2019distributions}, with moments:
$$\E(\lambda)=\mu;\qquad\Var(\lambda)=\mu^2\sigma.$$
The resultant distribution is bivariate negative binomial:
$$\begin{pmatrix}    
Y_1\\Y_2
\end{pmatrix}\sim \text{BivNB}(\mu_1, \mu_2, \sigma)$$
where $\mu_t=\mu t_t$, for $t=1,2$. The marginal distributions are \texttt{NBI}($\mu_t, \sigma$) (GAMLSS parameterization of the negative binomial) with moments 
\begin{align*}
    \E(Y_t)&=\mu_t \\
    \Var(Y_i)&=\mu_t+\sigma\mu_t^2 \qquad\text{ for } t=1,2,\\
    \Cov(Y_1,Y_2)&=\mu_1\mu_2\sigma.
\end{align*}

In this case $\lambda$ acts as a non-normal random effect term that mixes in a multiplicative way, as opposed to the additive way a random intercept is introduced in the standard mixed model. For this distribution, we generate 400 realizations with 1,000 observations at both time points for $\mu_1$ and $\mu_2$ between $5^{-3}$ and $5^3$ and $\sigma$ between $5^{-1}$ and $5$. 

\subsubsection{Bivariate Gamma}

We use the four-parameter bivariate Gamma of Nadarajah and Gupta\citep{Nadarajah2006}, which is described in Appendix \ref{sec:appendix biv gamma}. This distribution, which we denote as BivGamma$(\mu_1, \mu_2, \sigma, \theta)$, allows us to introduce significantly more skewed continuous distributions, and includes a unique dependence shape, which cannot be directly described by a parametric copula, nor generated by random effect. We align with the GAMLSS parameterization of the Gamma distribution \citep{McCullaghNelder89, rigby2019distributions}:
$$\begin{pmatrix}    
Y_1\\Y_2
\end{pmatrix}\sim \text{BivGamma}(\mu_1, \mu_2, \sigma, \theta)$$
 The marginal distributions are \texttt{GA}($\mu_t, \sigma$) with moments 
 \begin{align*}
 \E(Y_t)&=\mu_t\\
\Var(Y_t)&=\sigma^2\mu_t^2\qquad\text{ for } t=1,2,\\
\Corr(Y_1, Y_2)&=\frac{\sigma\sqrt{\theta}}{1+\sigma^2+\sigma^2\theta},
\end{align*}

where $\mu_t>0$ and $\sigma>0$ have the same meanings as in the GAMLSS parameterization \texttt{GA}; and $\theta>0$ is present only in the correlation.

\begin{figure*}[h]
    \centering
    \includegraphics[width=\linewidth]{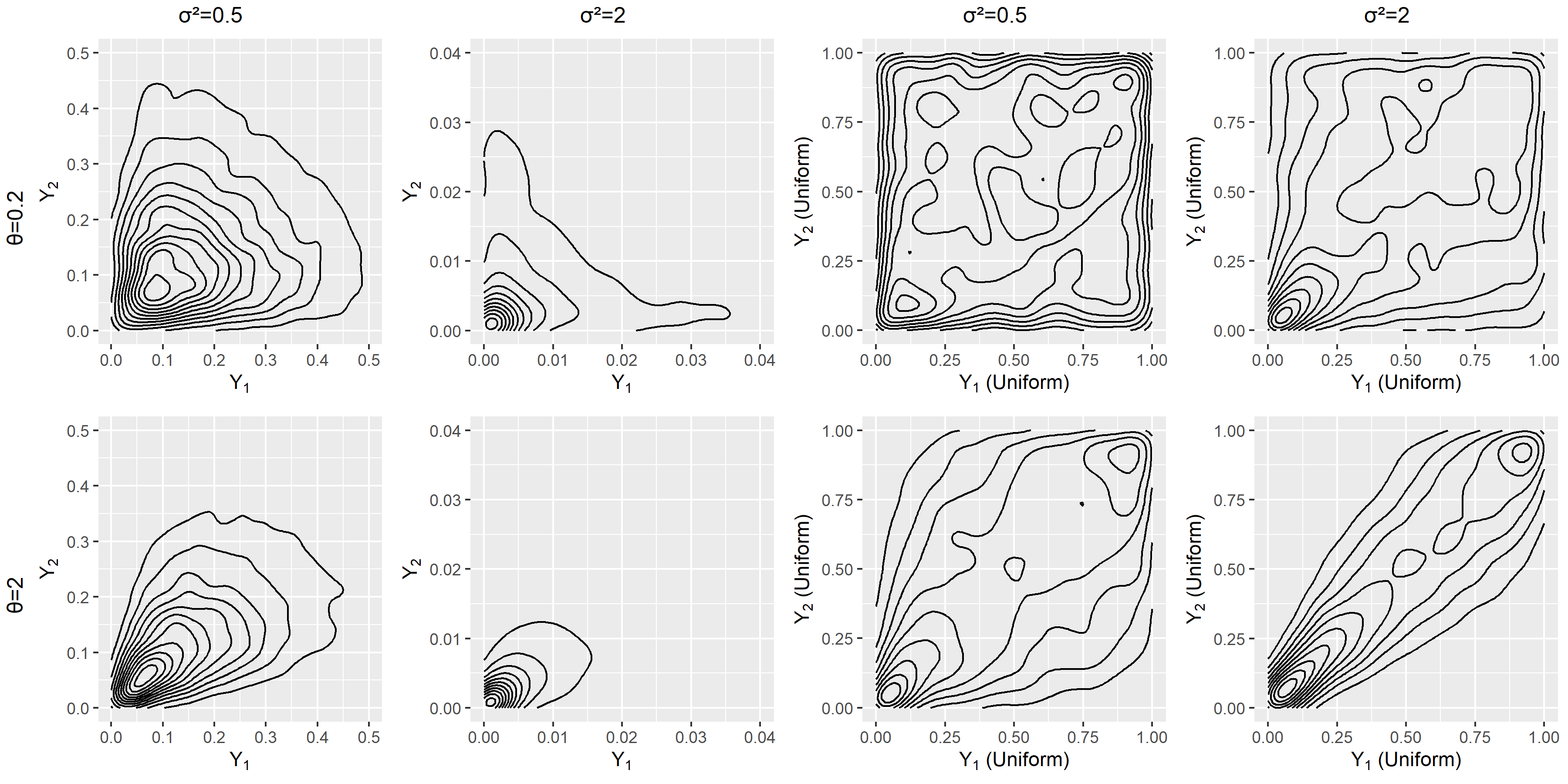}
    \caption{Shapes of the bivariate distribution of Nadarajah and Gupta for a simulation of 10,000 observations. Left four plots are contour plots of the bivariate distribution of $(Y_1,Y_2)$, for selected values of $\sigma$ and $\theta$, where $\mu_2=0.8\mu_1$ and $\mu_1=0.2$ when $\sigma^2=0.5$ and $\mu_1=0.05$ when $\sigma^2=2$. The right four charts are contour plots of the uniform transforms of the corresponding marginal distributions plotted against one another, to show the dependence structure with the marginal shapes removed.}
    \label{fig:bivariatesimcountours}
\end{figure*}

The shapes of the bivariate Gamma distribution and its dependence structure, for selected parameter values, are illustrated in Figure \ref{fig:bivariatesimcountours}. In Section \ref{sec:results}, the distribution is simulated with $n=1,000$ observations at both time points for 400 parameter combinations. The parameter settings chosen are: $\mu_1 \in (2,3,\ldots,21)$; $\mu_2=1.2\mu_1$; $\sigma \in(\sqrt{0.2},\sqrt{0.3},...,\sqrt{2.1})$; $\theta\in (0.2,0.3,\ldots,2.1)$. The resulting correlation varies between 0.08 and 0.4, and the rank correlation, as measured by Kendall's $\tau$, between 0.04 and 0.82. 
The distribution exhibits a higher marginal skew with higher values of $\sigma$ and a higher rank correlation with higher values of $\theta$, with the highest rank correlation when $\sigma$ is high and $\theta$ is high. 

\subsubsection{Bivariate Bernoulli}

The logistic model is extremely common in life sciences and allows us to introduce another link function to simulations: the logit link. We firstly generate simulations from a bivariate standard normal:
\begin{align*}
\begin{pmatrix}    
Z_1\\Z_2
\end{pmatrix}\sim\mathcal{N}_2\left(\begin{pmatrix}    
0\\0
\end{pmatrix},
\begin{pmatrix}
1&\rho\\  
\rho&1
\end{pmatrix}
\right)
\end{align*}

Then we transform the margins such that $Y_t=\mathbbm{1}(Z_t<C_t)$ where $C_t$ is defined as a set of different cut-off points to represent a success or failure for each observation corresponding to selected values for the mean at both time points: $\mu_1=\Phi(C_1)$ and $\mu_2=\Phi(C_2)$. This results in the bivariate Bernoulli distribution with Bernoulli marginals and the following moments:
\begin{align*}
    \E(Y_t)&=\mu_t \\
    \Var(Y_t)&=\mu_t(1-\mu_t) \qquad\text{ for } t=1,2,\\
    \Cov(Y_1,Y_2)&=\Prob(Z_1<C_1, Z_2<C_2)-\mu_1\mu_2
\end{align*}

The parameter settings chosen are: $\mu_1$ and $\mu_2$ in $(.1,.25,.5,.75,.9)$ so multiple combinations of different class balance could be tested, and correlation $\rho$ was tested in the range of values from $0.1$ to $0.9$.

\subsection{Model specification}\label{sec:model specification}

For the purpose of model specification, we assume that the bivariate data are longitudinal with $Y_1$ and $Y_2$ being the random variables for the response at times $t=1$ and $t=2$, respectively. For the sake of generality of estimation, we assume marginals $Y_t\sim\mathcal{D}(\mu_t, \sigma_t)$ where $\mathcal{D}$ is a distribution appropriate to the class of model (exponential family for GLMs and GEEs; any distribution with computable derivatives for GAMLSS, and a large subset of GAMLSS distributions that are available in GJRM); where $\mu_t$ is generally the mean; and $\sigma_t$ is a dispersion parameter. So then for the bivariate distributions we model the following distributional parameters:
\begin{itemize}
\item $\mu_1$ and $\mu_2$, the means of the marginal distributions,
\item $\sigma$, the dispersion of the marginal distributions, for the normal, Gamma and negative binomial, and
 \item $\theta$, the dependence parameter between the two marginals. This can be, for example, the Pearson correlation $\rho$, Kendall's $\tau$, or the copula parameter.
\end{itemize}

In addition we fit two covariates:
\begin{itemize}
    \item $x_1$, a binary factor with value either 0 or 1, with half of the individuals in each category, and
    \item $x_2$, a covariate uniformly distributed between 1 and 100.
\end{itemize}

We estimate the parameters of four classes of models that approximate the underlying bivariate data:
\begin{itemize}

\item \textbf{GLM} is the simplest model included as a benchmark that assumes independence between time points:
\begin{align*}
    Y_{it}&\sim\mathcal{D}(\mu_t,\sigma)\\
    g(\mu_{it})&=\beta_{1}\mathbbm{1}_{t=1} +\beta_{2}\mathbbm{1}_{t=2}+\beta_{x_1} x_1+\beta_{x_2} x_2\\
 \text{ for }&i=1,\ldots,n;\quad t=1,2.
\end{align*}
The model is specified as above so that $\beta_1$ and $\beta_2$ are the exclusive parameters for times 1 and 2 respectively. Note that the GLM has a constant dispersion parameter, so we have $\sigma_1=\sigma_2=\sigma$.

\item \textbf{GEE} incorporates an adjustment for dependence to the generalized linear model through a correlation matrix. In this case with only two time points there is only one correlation parameter, so there is no need to specify a correlation structure beyond this single parameter:
\begin{align*}
    Y_{it}&\sim\mathcal{D}(\mu_t,\sigma)\\
    g(\mu_{it})&=\beta_{1}\mathbbm{1}_{t=1} +\beta_{2}\mathbbm{1}_{t=2}+\beta_{x_1} x_1+\beta_{x_2} x_2 \\
    R_i=\theta &\text{ is the correlation parameter} \\
\text{ for }&i=1,\ldots,n;\quad t=1,2.
\end{align*}
We use the simplest form of the GEE, which follows the distributional specification of the GLM, i.e., a constant dispersion parameter $\sigma$.
 \item \textbf{GLMM} in its simplest form with just a single random intercept combines a fixed linear model with a normally distributed random intercept term, which is assumed to be common to subjects between margins and induces dependence between observations at the two time points.

For the GLMM, the five parameters of the bivariate distributions can be captured directly. The most flexible model is the generalized linear mixed model with a random effect term and time-dependent $\sigma$:
\begin{align*}
    Y_{it}&\sim\mathcal{D}(\mu_t,\sigma_t)\\
    g(\mu_{it})&=\beta_{1}\mathbbm{1}_{t=1} +\beta_{2}\mathbbm{1}_{t=2}+\beta_{x_1} x_1+\beta_{x_2} x_2+b_i\\
    g(\sigma_{t})& = \beta_{\sigma_{1}}\mathbbm{1}_{t=1} + \beta_{\sigma_{2}}\mathbbm{1}_{t=2}\\
    b_i&\sim\mathcal{N}(0, \theta^2)\\
    \text{ for }&i=1,\ldots,n;\quad t=1,2.
\end{align*}

We also consider the 4-parameter GLMM without a time-dependent dispersion parameter, i.e. where $g(\sigma_{t}) = \beta_{\sigma}$ for $t=1,2$; and an alternative model where the random effect term is specified as having a nonparametric rather than normal distribution.

\item \textbf{GJRM} jointly models the response variable at both times with a distributional fit, and its dependence structure through a copula function to capture the full bivariate distribution:
\begin{align*}
Y_{it}& \sim\mathcal{D}(\mu_t, \sigma_t) \\
g(\mu_{t})& = \beta_{1}\mathbbm{1}_{t=1} +\beta_{2}\mathbbm{1}_{t=2} +\beta_{x_1} x_1+\beta_{x_2} x_2\\
g(\sigma_{t})& = \beta_{\sigma_{1}}\mathbbm{1}_{t=1}+\beta_{\sigma_{2}}\mathbbm{1}_{t=1} \\
\theta &\text{ is the copula parameter} \\
\text{ for }&i=1,\ldots,n;\quad t=1,2.
\end{align*}

In the GJRM framework \citep{GJRM2023}, a distribution is selected for each of the margins and a copula function is selected to fit the dependence structure. For these simulated datasets, the marginal distributions are known. For the dependence structure, there are many varied copula shapes available for modeling. The Clayton, normal, Joe, Gumbel and Frank copulas are included in the simulations. 

For estimation we use the distribution which matches the marginals of the known bivariate distribution, e.g. Gamma marginal distribution for the Bivariate Gamma. We select the identity link function for all models for the normal, the $\log$ link function for the negative binomial and Gamma, and the logistic link function for the Bernoulli.

\end{itemize}

\subsection{Software}

Initially developed for normal response distributions, random effects for intercepts and covariate slopes have been incorporated in multiple distributional regression frameworks including generalized additive models, \citep[GAM][]{hastie1990generalized}, and generalized additive models for location, scale and shape, \citep[GAMLSS][]{Stasinopoulos2017}, making them a very accessible tool for modeling correlated data. GAMLSS provides a flexible regression framework that extends GAMs to allow multiple parameters of a response distribution to be modeled simultaneously. In addition, the range of potential response distributions is extended beyond the exponential family to any distribution with computable derivatives. Generalized estimating equations (GEE) have also been extended to apply to a much broader range of scenarios.

For the GJRM and GLM fits we use the \textbf{R} packages \pkg{GJRM} \citep{GJRM2023} and \pkg{glm} \citep{rstats_glm} respectively. For the GEE fits we use \pkg{glmtoolbox}\citep{vanegas_generalized_2023} which provides a modern implementation of the GEE with significant additional features for diagnostics and fitting. 

For GLMMs, there are multiple packages available for fitting. In these simulations, we utilize four packages that use different methods. We include methods primarily based on their popularity and ability to capture the various different approaches used for optimization of the fit. This enables us to compare not just GLMMs to alternative models, but also compare the implementations of the GLMM as the methods for model fitting significantly differ. The methods for GLMM that are included are:
\begin{itemize}
    \item 
The package \pkg{gamlss} \citep{gamlsspack} is included for fits of four and five parameters with normally distributed random effects. It utilizes the \texttt{random()} function to fit random effects and optimizes based on a penalized likelihood function; 
\item \pkg{lme4} \citep{lme4pack} is included as it is one of the most popular packages for fitting generalized linear mixed models with random effects, and uses a Laplace approximation to the likelihood function; 
\item The function \texttt{gamm()} from \pkg{mgcv}\citep{wood2001mgcv} is included for fits of four parameters with normally distributed random effects. This utilizes \texttt{glmmPQL()} to optimize a penalised quasi-likelihood function; 
\item \pkg{gamlss.mx}\citep{gamlss.mx}, in the \pkg{gamlss} suite of packages, is included for the fitting of a non-parametric random effect term, in a model with five parameters. 
\end{itemize}
We refer to these methods in shorthand as GAMLSS (4), GAMLSS (5), LME4, GAMM, and GAMLSS NP, respectively. Throughout our simulations, we found that the results from the GAMLSS (5) model were highly unstable, providing extreme results at a high rate. These have been excluded from the presented results.

All code used for this paper is available in a public repository at \url{https://github.com/ahibbert/bivariate-copula-for-correlated-data}.

\section{Results}
\label{sec:results}

Our evaluation of model performance covers three areas to provide a broad view of the difference in model fits across the methods presented, and to provide the ability to highlight areas of strength or weakness for the different models.

First, we evaluate estimated model \textbf{coefficients} and \textbf{standard errors} against the true values. This provides a view as to the models' suitability for inference, both in terms of the estimate and the standard error that would be used in hypothesis testing. Models which provide closer coefficient estimates to the true value on average, or provide more consistenly close estimates are preferred. Similarly, providing accurate standard errors for parameters on average, or more consistently close estimates to the true values are preferred.

Second, we evaluate model overall fit on the basis of two \textbf{model selection criteria}. We use multiple criteria due to advantages and drawbacks of each approach. 

One criterion used is Bayesian Information Criteria (BIC), defined as $p\ln(n)-2\ell$, where $\ell$ is the log-likelihood of the model fit, $p$ is the number of parameters and $n$ is the sample size. BIC generally provides the more parsimonious model for inference compared with AIC, which provides the best model for prediction\citep{stasinopoulos_generalized_2024}. For reference we also present the log-likelihood in the main text, and AIC and GAIC (4) (AIC with a penalty for $p$ of 4) in Appendix \ref{sec:appendix_aic}. Some adjustment to the likelihood criteria is needed: for the GLMM models, conditional AIC \citep{donohue_conditional_2011} (cAIC) and its extension to BIC has been used to take into account the variability of random effects on the degrees of freedom of the model. In the GAMLSS GLMM optimization method, the degrees of freedom are estimated by the degrees of freedom of the smoother needed to fit the random effect. This effective degrees of freedom (EDF) is calculated directly by the GAMLSS packages, and is roughly comparable to the same estimate for EDF for a GLMM in the calculation of conditional AIC (cAIC) used for the AIC and BIC estimates for the LME4 models. The GAMLSS, GAMLSS NP and LME4 methods all provide valid likelihoods and so are adjusted to cAIC. Unfortunately, while this measure is highly sensitive, it is not available for the GEE as this uses quasi-likelihood for optimization; or GAMM as its optimization is based on penalized quasi-likelihood. Hence we require an additional method to enable comparison of all models. The use of a method which does not score on the training data would be ideal to protect against overfitted models. We also note that while quasi-likelihood criteria are available for the GEE, they are only comparable between GEE models.

The other model selection criterion used is the Variogram Score\citep{gneiting_strictly_2007} with $p=2$. This is an extension to the concept of the Energy score, which provides significantly more sensitivity to incorrectly specified correlations, of interest for this model comparison. Variogram scores provide a pairwise comparison of all datapoints between a simulated fitted model and simulated realizations of the true dataset, essentially comparing the true distribution to the one proposed by the model. The value of $p$ is the power of the differences, so $p=2$ provides squared pairwise differences. As the Variogram Score does not rely solely on the realized dataset being fit, the score provides a criterion which is more robust against overfitted models than likelihood-based criteria. Variogram scores are shown to be sensitive to incorrectly predicted means, variances and correlations\citep{noauthor_variogram-based_nodate}, allowing us to assess the full distributional fit in a similar way to the log-likelihood. 

Cross-validated Mean Squared Error of Prediction and Log Score were both considered as alternative approaches for assessment of overall fit. However, we lack a prediction method for the GJRM which can incorporate the effect of correlation for a time holdout, which would be required for an accurate out of sample assessment. We also lack the ability to predict for an unobserved individual in many models, which would be needed for an in-time-point holdout. Hence these methods are unable to provide a comparison between all methods and so are not included.

Third, we evaluate model \textbf{computational complexity} through runtime and convergence of each model across the range of simulations.

\subsection{Coefficient estimates and standard errors}\label{sec:mean_model_results} \label{sec:marginal_parameters}

Each of the models specified in Section \ref{sec:model specification} is applied to the simulated datasets. Estimates for parameters and their standard errors are compared to their true values under the simulated bivariate distribution. 

The key results for the simulations are shown in Figure \ref{fig:simBias} for the intercepts and Figure \ref{fig:simBiasX1} for the covariate coefficients $\beta_{x_1},\beta_{x_2}$. As the bias and variance of the estimates appear to be highly dependent on Kendall's $\tau$, we present these as a function of $\tau$.

\textbf{Intercept coefficients}

Bias and standard error results for intercepts for times 1 and 2 are extremely similar, so for simplicity only results for time 1 are shown; results for time 2 are available in Appendix \ref{sec:appendix_coefficients}. Figure \ref{fig:simBias} provides the average bias (left plots) and standard error (right plots) for time 1 intercept, for each of the models, across Kendall's $\tau$ of the simulated distributions. The black lines indicate the true asymptotic values for the bias (horizontal line at zero) and standard error. The true values of the parameters and their asymptotic standard errors are calculated based on simulation of the joint distributions in Section \ref{sec:bivariate_distributions}, and the asymptotic properties of the maximum likelihood estimators.

\begin{figure*}[htbp]
 \includegraphics[width=\linewidth]{Charts/Mu1_Bias_and_SE_vs_Correlation_Combined_2025-10-23.png}
 \caption{Bias for the estimate of the intercept for time 1 plotted with inverse link transform $g^{-1}({\beta_1})$ (left) and standard error for $\beta_1$ (right), across the range of realizations of the simulated distributions. Estimates are plotted against Kendall's $\tau$ which has been computed for each simulation. Random effect models (GLMMs) are shown as dashed lines, GJRMs are solid and GLM, GEE are dotted lines for ease of reference. 
 }
 \label{fig:simBias} 
\end{figure*}

Beginning with the \textbf{bivariate normal} (top row), the results show that all models exhibit minimal bias throughout the range of simulated distributions, except for the GJRM with an ill-fitting skewed copula (Clayton), which is included for comparison. The GJRMs, GEE and GLM provide close estimates for the standard error (SE); however, while LME4 and GAMM provide broadly accurate standard errors, the GAMLSS (4) model shows an incorrectly decreasing standard error for realizations with higher $\tau$.

In contrast to the normal model, for the \textbf{non-normal models}: the bivariate negative binomial (second row), bivariate Gamma (third row) and bivariate Bernoulli (bottom row), we find substantial bias in the parameter estimates for the intercept from the parametric GLMMs (LME4, GAMLSS(4), GAMM) while other models are relatively unbiased. In investigating the characteristics of simulated distributions that result in biased parametric GLMM estimates, our finding is that higher values of Kendall's $\tau$ and / or higher values of skewness of the marginal distributions of the simulated bivariate distribution result in higher bias. In addition, there are some large differences in estimates for parameters and standard errors between GLMM implementations. 

\begin{figure*}[h!]
\includegraphics[width=\linewidth]{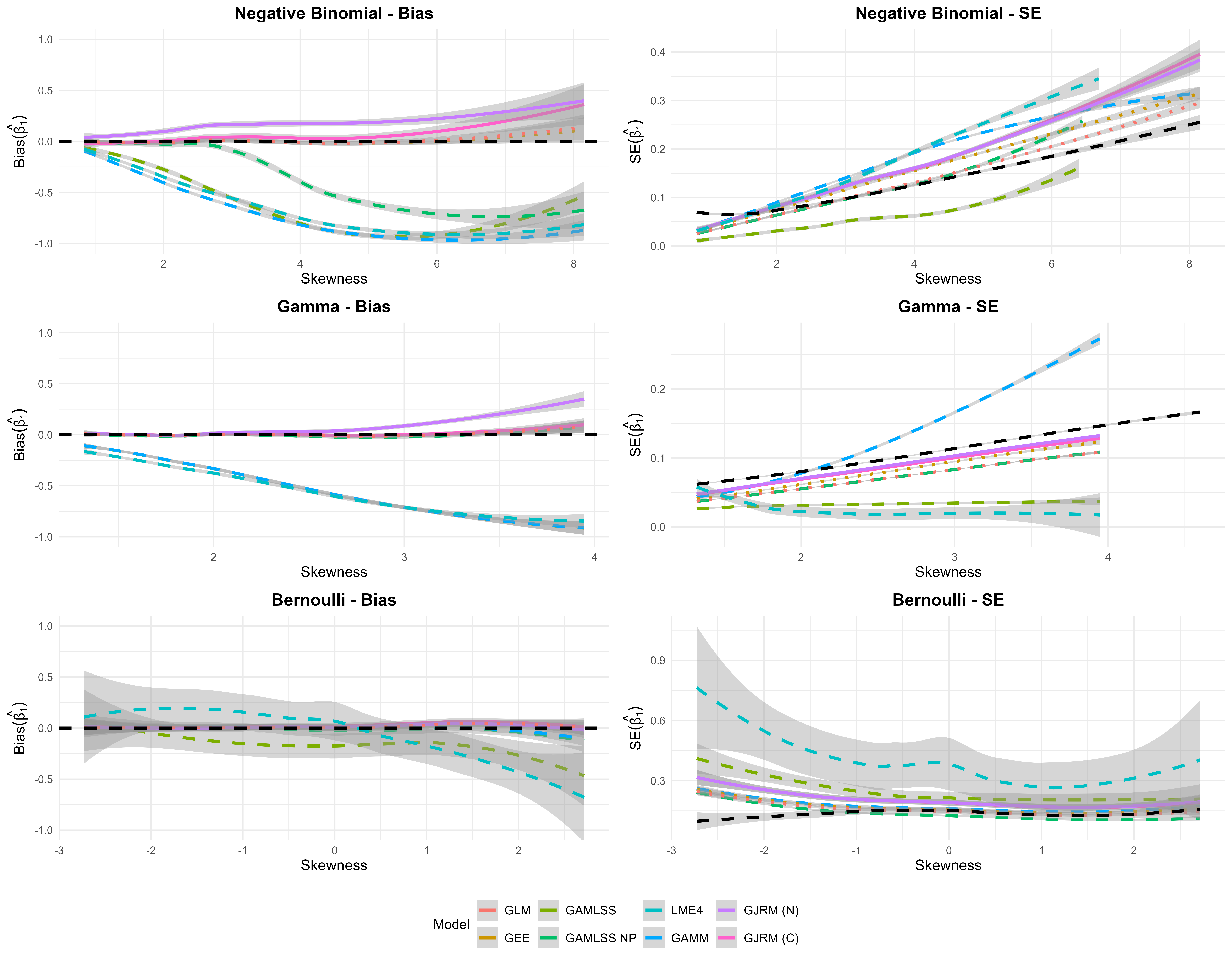}
 \caption{Bias for the estimate of the intercept for time 1 plotted with inverse link transform $g^{-1}({\beta_1})$ (left) across the range of realizations of the simulated distributions, plotted against average marginal skewness. Random effect models (GLMMs) are shown as dashed lines, GJRMs are solid and GLM, GEE are dotted lines for ease of reference.}
 \label{fig:simBiasSkew} 
\end{figure*}

For the \textbf{negative binomial}, bias for all four GLMM estimates (GAMLSS (4), LME4, GAMLSS NP (5), GAMM) is exhibited reasonably consistently across the range of Kendall's $\tau$ values, with the remaining models being relatively unbiased. For this distribution, skewness is a stronger indicator of GLMM bias. The bias against the marginal skew is provided in Figure \ref{fig:simBiasSkew}. Standard error estimates for the negative binomial parameters are close to the true values for the GJRM, GEE, and GLM. The GLMMs differ substantially for estimates of the standard error, with the GAMLSS (4) model underestimating SE for higher $\tau$, and LME4 and GAMM slightly overestimating SE across the range of $\tau$. 

In the case of the \textbf{bivariate Gamma}, bias for the parametric GLMMs increases almost linearly with increasing $\tau$ for the GAMLSS (4) model, GAMM and LME4. All other models provide unbiased estimates for the intercept except for the GJRM with a normal copula that is biased for higher $\tau$, due to its ill-fitting copula. In extreme cases with rank correlation greater than 0.6, the GLMM estimates from the parametric GLMMs are less than half the simulated intercept. Interestingly, we also found that the LME4 and GAMLSS (4) parametric GLMMs provide much more inconsistent estimates for the intercept in the simulation range, with a much greater variation in estimate bias across values of $\tau$ compared to the other models. In comparison, the GLM, with a similar underlying structure but without adjustment for dependence, is robust, providing unbiased estimates for the intercept. 

In terms of standard error for the Gamma, the GLM, GEE and GJRMs closely follow the true standard error. In contrast, the GLMM implementations again exhibit differing behaviour. The GAMLSS (4) and LME4 models provide a generally flat standard error with increasing values of $\tau$, severely underestimating the true standard error, and the GAMM model follows the trend of the true standard error, but is overestimating the true standard error substantially for high $\tau$. 

Model bias for the intercept plotted against skewness for all non-normal models is shown in Figure \ref{fig:simBiasSkew}. The GLMMs provide increasingly biased estimates for higher values of marginal skewness, whereas the remaining models remain relatively unbiased throughout the range. 

Across the set of realizations, we find that for the Gamma, increasing marginal skewness holding $\tau$ constant, or increasing $\tau$ and holding skewness constant, results in increasing bias for the GLMMs. This is shown in Table \ref{tab:bias_mu1_side_by_side} (LHS) for the non-time-variant-$\sigma$ random effect model, GAMLSS (4). This is also the case for the GJRM with less-well fitting symmetric normal copula being fit to the skewed dependence of the bivariate Gamma (Table \ref{tab:bias_mu1_side_by_side} (RHS)).

\begin{table}[ht]
\centering
\label{tab:bias_mu1_side_by_side}
\setlength{\tabcolsep}{3.5pt}

\begin{subtable}[t]{0.49\textwidth}
\centering
\small
\begin{tabular}{l *{4}{S[table-format=-1.3]}}
\toprule
& \multicolumn{4}{c}{Marginal skewness for $Y_1$} \\
\cmidrule(lr){2-5}
{$\tau$} & {1--2} & {2--3} & {3--4} & {4--5} \\
\midrule
0.0--0.2 & -0.099 & -0.262 & \NA & \NA \\
0.2--0.4 & -0.233 & -0.424 & -0.722 & -0.942 \\
0.4--0.6 & -0.318 & -0.541 & -0.822 & -0.961 \\
0.6--0.8 & \NA & -0.692 & -0.853 & -0.972 \\
0.8--1.0 & \NA & \NA & \NA & -0.976 \\
\bottomrule
\end{tabular}
\caption*{GAMLSS (4)}
\end{subtable}
\hfill
\begin{subtable}[t]{0.49\textwidth}
\centering
\small
\begin{tabular}{l *{4}{S[table-format=-1.3]}}
\toprule
& \multicolumn{4}{c}{Marginal skewness for $Y_1$} \\
\cmidrule(lr){2-5}
{$\tau$} & {1--2} & {2--3} & {3--4} & {4--5} \\
\midrule
0.0--0.2 & -0.001 & -0.006 & \NA & \NA \\
0.2--0.4 & -0.024 & -0.005 &  0.102 &  0.269 \\
0.4--0.6 &  0.002 & -0.016 &  0.105 &  0.304 \\
0.6--0.8 & \NA &  0.023 &  0.215 &  0.782 \\
0.8--1.0 & \NA & \NA & \NA &  1.050 \\
\bottomrule
\end{tabular}
\caption*{GJRM (N)}
\end{subtable}
\caption{Average bias for the estimate of $\beta_1$ presented in inverse link transform $g^{-1}(\beta_1)$ against $\tau$ and marginal skewness for realizations of the bivariate Gamma. Left hand side table is for the GAMLSS (4) model and right hand side table is for the GJRM with normal copula, which is ill-fitting for this distribution compared to the Clayton copula.}
\label{tab:bias_mu1_side_by_side}
\end{table}

For the \textbf{bivariate Bernoulli} using a logit link (logistic model), results are similar to that of the negative binomial and Gamma models. The GJRM, GEE and GLM all provide reasonable estimates of the marginal intercepts and their standard errors. However, for the GLMMs, bias presents for the marginal intercept estimates across the range of $\tau$ with higher bias for higher $\tau$. However, while the GAMM estimates are biased, its estimates are much closer to the true values than the other parametric GLMMs, being only 5 percent different than the true value at the worst point compared to 20-30 percent for the other GLMMs. Again, for standard error, the GLMMs differ substantially, with the LME4 model providing extremely high standard errors, compared to the GAMLSS (4) model which is only slightly higher than the known true values, while the GAMM provides an accurate estimate in line with the other models and the true value.

In summary, for the bivariate normal, all models provide reasonable estimates of  the intercept and its standard error for the specified parameters, except for the GJRM where the copula is ill-fitting. For the negative binomial, Gamma, and Bernoulli, the GLM and GEE provide consistently unbiased estimates of the intercept and standard error. The GJRM similarly provides unbiased estimates for the intercept and standard error across all distributions as long as the copula function is well specified. When the copula function is ill-fitting, bias presents for high $\tau$ and high marginal skewness, similar to the GLMMs. For the GLMMs, the results differ substantially between packages. In general, we find that for the non-identity link models, the GLMMs provide increasingly biased estimates of the intercept with higher values of $\tau$ and marginal skewness. The GLMMs also often significantly under- or over-estimate the standard error with different trends depending on the package used, and inconsistency is higher with higher $\tau$ and marginal skew.

\textbf{Coefficients} $\bm{\beta_{x_1}, \beta_{x_2}}$

Results for coefficients $\beta_{x_1}$ and $\beta_{x_2}$ are very similar so we only include results for $\beta_{x_1}$ in Figure \ref{fig:simBiasX1} and include $\beta_{x_2} $ in Appendix \ref{sec:appendix_coefficients} for reference.
For the normal, all models are reasonably accurate for $\beta_{x_1} $ and its standard error, except for GAMLSS which underestimates SE for high correlation, as was the case for $\beta_1$.

Interestingly, for the negative binomial, the GJRM with ill-fitting copula provides the most biased estimates, particularly for high correlation, while the GJRM with well-fitting copula, and all other models, estimates are reasonable. However, the GAMLSS model also underestimates SE across the range for the negative binomial, while other GLMMs are reasonably accurate, though slightly higher than the true value and other models.

For the Gamma, all models are reasonably unbiased for $\beta_{x_1}$ except for regions of very high correlation. At above $\tau=0.6$, almost all models except the GJRM with normal copula underestimate the parameter by about 5 percent. SE is accurately captured by the GJRMs, GLM and GEE but is highly underestimated by the LME4 and GAMLSS models, while being highly overestimated by the GAMM. The only substantial difference for the estimation of $\beta_{x_2}$ compared to that of $\beta_{x_1}$ is that for the Gamma for high $\tau$, the GLMMs overestimate $\beta_{x_2}$, as opposed to for $\beta_{x_1}$ where they underestimate the coefficient.

Estimates for $\beta_{x_1}$ for the Bernoulli are highly unusual. The GAMLSS and LME4 models provide highly biased over-estimates for $\beta_{x_1}$, but strangely the two models differ on SE estimates: the GAMLSS model heavily under-estimates the SE and the LME4 model heavily over-estimates it.

In summary, all models provide relatively reasonable and comparable estimates for the coefficients for the normal case. For the other distributions, compared to estimates of the intercept, the bias for coefficient estimates from the GLMMs is substantially lessened for $\beta_{x_1}$ and $\beta_{x_2}$ but still present for high correlation and skewness for the negative binomial and Gamma, and still highly present for the Bernoulli. In general, SE results for $\beta_{x_1}$ and $\beta_{x_2}$ are inconsistent between GLMM packages. The \pkg{gamlss} \texttt{random()} implementation consistently underestimates the SE of the coefficient to the extent that it would substantially change interpretation; \pkg{lme4} overestimates SE substantially for the Bernoulli and underestimates substantially for the Gamma; and \pkg{mgcv} \texttt{gamm()} software over-estimates for the Gamma and is reasonable for the Bernoulli.

\begin{figure*}[h!]
\includegraphics[width=\linewidth]{Charts/X1_Bias_and_SE_vs_Correlation_Combined_2025-10-21.png}
 \caption{Bias for the estimate of the coefficient for $x_1$ across the range of realizations of the simulated distributions, plotted against Kendall's $\tau$. Random effect models (GLMMs) are shown as dashed lines, GJRMs are solid and GLM, GEE are dotted lines for ease of reference.}
 \label{fig:simBiasX1} 
\end{figure*}

\subsubsection{Reviewing the differences in coefficient estimates between GLMM and other models}\label{sec:differencereview}

The significant difference in bias for the GLMMs between the normal distribution models and the negative binomial and Gamma distribution models provides an indicator of the mechanism that drives the bias. There are three key differences between the normal and other distribution models: (i) most obviously, the marginal distributions; (ii) the shape of the dependence structure; and (iii) the link function.
The results of simulations in this paper indicate that if using a non-identity link function with a distribution with significant skewness and/or correlation, then inferences based on conditional parameters may be wildly different from their marginal values. Essentially, the higher the correlation (measured by Kendall's $\tau$) and the higher the marginal skewness, the greater distortion the random effect term will cause to the conditional distribution in comparison to the true marginal distribution.

Although unintuitive, this bias is expected behavior for the GLMM coefficient estimates to some extent based on the maximum likelihood estimators. To illustrate the impact of the link function on the estimator, we introduce a new parameter $\beta_t=\beta_2-\beta_1$. In Appendix \ref{sec:appendix_MLE} we show derivations for the exponential family, for the maximum likelihood estimators for $\beta_1$, $\beta_2$ and $\beta_t$ with the log link, for a marginal and conditional model. We use these estimators to show why bias arises for parameters $\beta_1$ and $\beta_2$ but not $\beta_t$. For simplicity we don't include covariates $x_1$ and $x_2$ in this comparison but the working extends to these.

For the parameters $\beta_1$ and $\beta_2$ we have the following maximum likelihood estimators.

For the marginal model:
$$
\hat{\beta_1}=\log(\overline{y}_1 ); \qquad 
\hat{\beta_2}=\log(\overline{y}_2 )
$$
For the conditional model:
$$   \hat{\beta_1}=\log(\overline{y}_1 ) - \log\left(   \textstyle\sum_{i=1}^n e^{b_i}\right); \qquad
    \hat{\beta_2}=\log(\overline{y}_2 ) - \log\left(   \textstyle\sum_{i=1}^n e^{b_i}\right). 
$$

So the value for $\hat{\beta_1}$ and $\hat{\beta_2}$ will differ between the conditional GLMMs and marginal models based on the scale of $\sum{e^{b_i}}$ relative to $\sum y_{it}$.

Interestingly, for the logarithmic link function, while the estimates of the intercept cannot be easily extracted from the conditional model as $\E(Y_t)\neq\E(Y_t|b)$, the model can be respecified so that we can extract one of the parameters without a difference between the conditional and unconditional (marginal) model estimates. Consider the model:
\begin{equation} \label{eq:1}
    \log(\mu_{it})=\beta_1+\beta_t\mathbbm{1}_{t=2} + b_i
\end{equation}

In Appendix \ref{sec:appendix_MLE} we show that the MLEs for $\beta_t$ for both marginal and conditional models are:
$$
\hat{\beta_t}=\log(\overline{y}_2 )-\log(\overline{y}_1 ).
$$
So then the conditional and marginal estimates for $\beta_t$ using the log link are equivalent, due to the cancellation of $\sum b_i$ terms. As confirmation, we have rerun all the models in previous sections with the model respecified as Equation \eqref{eq:1}. We find that the results for the first parameter $\beta_1$ are consistent with the results presented for the GLMM in previous sections (biased and incorrectly estimating SE), while the time parameter $\beta_t$ is unbiased (though still presents similar inconsistencies in standard error estimates as for other parameters from GLMM fits). 

This specific simplification for the estimation of the time parameter, $\beta_t$ in this model, is the only parameter we have shown to be unbiased for non-normal data with logarithmic link and is only valid for a longitudinal factor such as the effect of time in this model. For example, the covariates that are constant over time, such as age at the start of the study or sex (e.g. $x_1$ and $x_2$), the factor would not be able to be extracted independently of $b_i$ and will not provide unbiased estimates. We are not aware of other forms of parameter formulation that may be defined without bias presenting. 

More generally, only a covariate that is specified such that its value is independent of $b_i$ when using a given non-identity link function can be extracted without bias from the GLMM for a marginal parameter value. This form will differ depending on the link function.

\subsection{Model selection criteria}

\begin{figure*}[h]
\includegraphics[width=\linewidth]{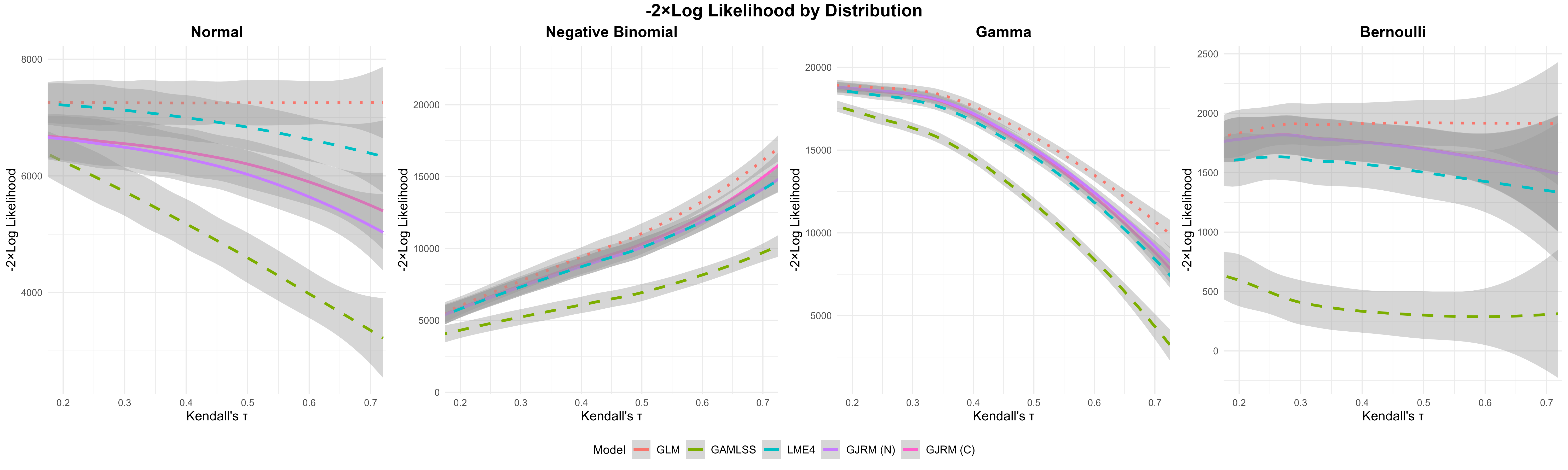}
 \caption{Average model selection criteria (-2$\times$log-likelihood) for the likelihood-based models across the range of realizations of the simulated distributions. Estimates are plotted against Kendall's $\tau$ which has been computed for each simulation. Random effect models (GLMMs) are shown as dashed lines, GJRMs are solid and GLM, GEE are dotted lines, for ease of reference.}
 \label{fig:loglik} 
\end{figure*}

\begin{figure*}[h]
\includegraphics[width=\linewidth]{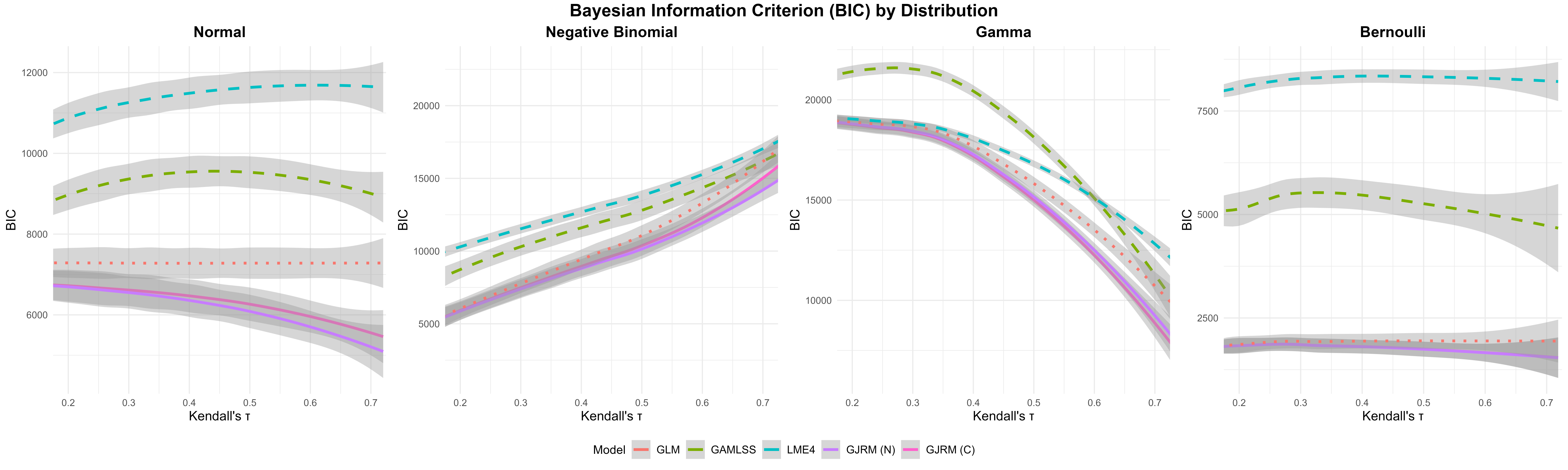}
 \caption{Average model selection criteria (BIC) for the likelihood-based models across the range of realizations of the simulated distributions. Estimates are plotted against Kendall's $\tau$ which has been computed for each simulation.}
 \label{fig:bic} 
\end{figure*}

The averages of the model selection criteria against $\tau$ are presented for all simulations. Logarithmic likelihood is shown in Figure \ref{fig:loglik}, Bayesian Information Criteria (BIC)\citep{stasinopoulos_generalized_2024} in Figure \ref{fig:bic}, and weighted variogram scores ($p=2$) in Figure \ref{fig:vs2}. Note that we present -2$\times$log-likelihood instead of likelihood so that the three plots can be interpreted as the best model having the lowest result. Results for AIC and generalized AIC with penalty $k=4$ (denoted GAIC (4)) are provided in Appendix \ref{sec:appendix_aic}. The results for AIC and GAIC(4) are broadly comparable between the GJRM and GAMLSS (4) model and so are not presented in detail.

The GEE does not provide a valid log-likelihood and therefore is not included in likelihood-based comparisons. This is a significant advantage of the GJRM method over the GEE, as it can be compared to alternative models such as the GLM and GLMM using standard likelihood-based criteria. For the variogram score we are able to simulate all models except the non-parametric random effect model, GAMLSS NP. 

GAMM provided identical log-likelihood results to LME4 for the normal, but provided extremely unusual likelihood results in comparison to other GLMMs for non-normal models, for example, five times lower log-likelihood for the logistic model, and so are excluded from comparison and require further investigation. 

\subsubsection{Likelihood-based model selection criteria}

The findings across all models are quite consistent for likelihood-based criteria. The GAMLSS random effect model with 4 parameters (green dashed line) is consistently chosen as having the lowest (best) -2$\times$log-likelihood across all distributions, while also having one of the two highest (worst) BIC alongside the LME4 model. The GJRMs provide the next-best log-likelihoods across most distributions after the GAMLSS model while providing the lowest (best) BIC on average across all bivariate distributions. For the negative binomial and Gamma, the LME4 and GJRM models have comparable likelihood, while the GJRM is preferred by likelihood for the normal and LME4 model is preferred by likelihood for the Bernoulli.

The GLM and GAMLSS GLMM with a non-parametric random effect are rarely preferred to the GJRM or parametric GLMMs. Similarly, the LME4 model is rarely preferred to the GAMLSS GLMM. In terms of copula choice for the GJRM, as would be expected, the GJRM with normal copula is preferred for the multivariate normal and the GJRM with Clayton copula is preferred for the skewed dependence structure of the bivariate Gamma though only marginally, while being very similar for the remaining models. Broadly both copulas seem to provide similar fit.

The high BIC results for the GLMMs is due to the high effective degrees of freedom (EDF) needed to account for the variability of the random effect used for those models. This means that model choice using standard model selection criteria between GAMLSS (4) and the GJRM is largely dependent on the penalty for the degrees of freedom of the random effect. The choice is between the more parsimonious model of the GJRM selected by BIC or the more closely fitting GLMM selected by log-likelihood.

\subsubsection{Variogram-based model selection criteria}

In addition to likelihood-based model selection criteria, we also compare model fit using variogram scores with $p=2$, taking the average of 100 variogram calculations for each model fit for each distribution and combination of parameters. 

\begin{figure*}[!htbp]
\includegraphics[width=\linewidth]{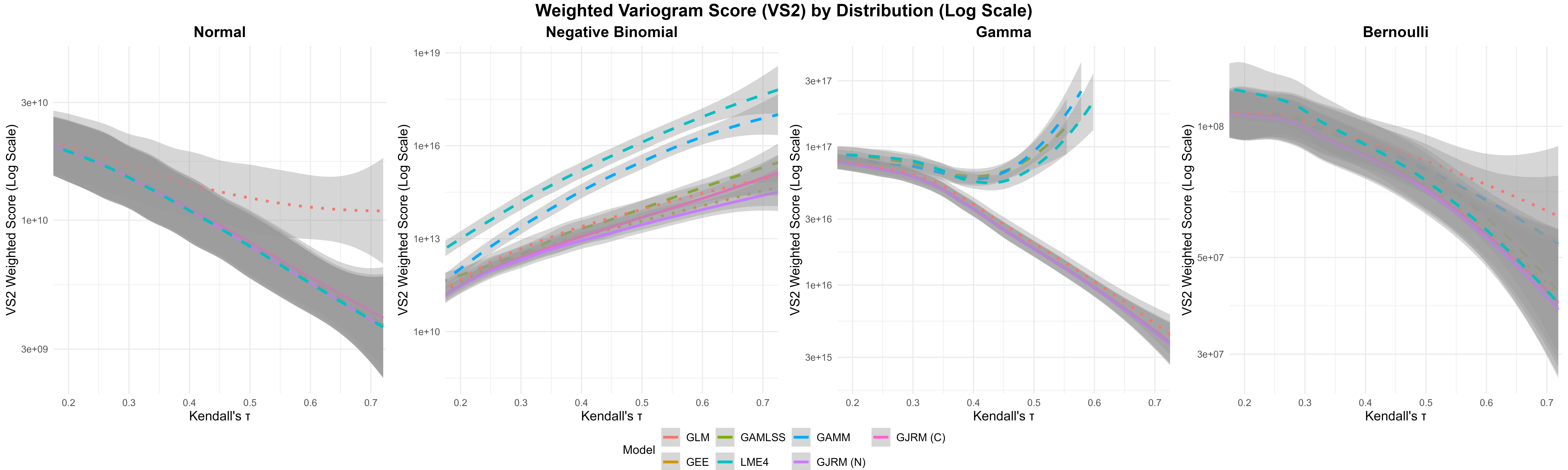}
 \caption{Average model selection criteria (weighted variogram score) for the models we can simulate (excluding non-parametric random effect GAMLSS NP) across the range of realizations of the distributions. Estimates are plotted against Kendall's $\tau$ which has been computed for each simulation.}
 \label{fig:vs2} 
\end{figure*}

Our initial findings were that the variogram score without weighting provided minimal differentiation between models for the normal, failing to differentiate even the GLM compared to correlation adjusting models at a high correlation. This is due to the nature of the bivariate data, whereby correlated observations only account for $2(n-1)/n^2$ of the score, and as they are correlated contribute less to the overall differences in the full pairwise matrix used to calculate the variogram score. To account for this problem in the score, we apply a weighting to the variogram calculation, which assigns a different weight to the subset of correlated observations in the dataset, while setting the weight for all other observations to the default. We find a weighting of $[{[n^2-2(n-1)]/[2(n-1)]}]^2$ for the correlated observations provides a reasonable balance between assessing the marginal fit and the correlation structure. This differentiation can be seen well in the first plot of Figure \ref{fig:vs2} which is for the bivariate normal, where all models provide almost identical fits excepting the GLM (orange dotted line) which is being differentiated as a worse fit (higher variogram score) with increasing correlation of the modelled dataset. Results for the unweighted variogram score ($p=2$) are included in Appendix \ref{sec:appendix_aic}.

For the negative binomial and Gamma which both use the log link, we find that the random effect-based models (dashed lines) all provide increasingly worse fits compared to the other models with increasing correlation of the underlying dataset. A review of these situations finds that the issue is with the increasing contribution of the simulated random effect to the overall fit and its interaction with the link function. With larger random effects relative to the marginal model, simulated results in the tail of the random effect distribution are exponentiated, and result in more highly skewed simulated distributions in comparison to the true marginal distribution. This may be the same issue that contributes to convergence issues in fitting GLMMs to these highly skewed datasets. In contrast, for the logisitic model, this issue does not occur in simulations of the random effects as the link function controls the impact of the random effect on the marginal distribution more effectively. This link function effect results in similar fit performance for all models across the range of correlation for the Bernoulli, excepting the GLM, which provides worse fits due to not accounting for correlation as would be expected. 

Surprisingly, while the weighted variogram score is able to differentiate the GLMM models when they provide extreme values for the log link distributions, and the GLM models which are ignoring correlation, the score does not seem to provide substantial differentiation for the more subtle differences between models which are correctly adjusting for correlation. For the Gamma and negative binomial, the GJRMs do in fact provide the lowest average weighted variogram scores, very closely followed by the GEE, however the error bands are very broad, and it is reasonable to conclude that all correlation-adjusting models are highly comparable when compared by weighted variogram, excepting for GLMMs with a log link for high-correlation, high-skew distributions.

\subsection{Computational complexity}

The computational complexity of each method is a significant potential factor influencing choice of model. With increasing complexity in model structure such as the GLMM and GJRM over the GLM, comes increased time to compute.

To compare models, we provide the runtime results for each of the models over 100 simulations of two different parameter combinations at different ends of the correlation range, and at different sample sizes. Results are presented in Table \ref{tab:runtime_comparison}. We find no significant differences in runtime for different parameter settings of the same distribution for a given model so simply provide timing results by distribution and sample size. For simplicity we only present results for two of the GJRM models, with normal and Clayton copulas, as runtimes between copulas is extremely similar.

At a sample size of $n=2000$, the order of model computation speed is generally consistent. In all cases the GLM provides the fastest computation, albeit without accounting for correlation, followed by the GEE taking roughly ten times longer than the comparable GLM. Of the remaining models, the GJRMs are the fastest, averaging two seconds per run, substantially faster than the 4 seconds for GAMLSS NP, 6 seconds for GAMM and GAMLSS, and 9 seconds for LME4. The one exception is that for the normal, LME4 seems to be highly optimized and is in fact faster than the GEE just for this case. Conversely, LME4 has by far the longest runtimes for the negative binomial models, and in fact is the only model which presents any convergence errors throughout these simulations. At $n=2000$, the LME4 model provided convergence warnings for 70 per cent of negative binomial fits and 97 per cent of Gamma fits even with an extremely high iteration limit. At $n=200$, LME4 provided much more stable results with only 1 percent of models failing for the negative binomial and 69 per cent for the Gamma. In contrast, the GAMLSS and GAMM models did not provide any convergence warnings for these fits, though it may be possible that the unusually low standard error results for the model across simulations may be caused by convergence errors which are not being identified in the optimization approach.

We provide a ratio of the total runtime between the fits at $n=200$ and $n=2000$ to provide an indication of how model runtimes scale with sample size. A model that scales linearly with sample size would provide a ratio of 10x for this case. We find that the GLM provides the most favourable scaling with sample size at 3.6x, followed by the GJRMs which average 5.3x, followed by the remaining models which are all above 7x. The GAMM and GAMLSS NP models in particular seem to scale the most unfavourably with this change in sample size.

\begin{table}[htbp]
\centering
\begin{tabular}{llrrrrrrrr}
\toprule
\multirow{2}{*}{Sample} & \multirow{2}{*}{Runtime} & \multicolumn{8}{c}{Models} \\
\cmidrule(lr){3-10}
& & GLM & GEE & GAMLSS & GAMLSS NP & LME4 & GAMM & GJRM (C) & GJRM (N) \\
\midrule
\multirow{5}{*}{n=200} & Gamma & 0.006 & 0.04 & 0.42 & 0.34 & 0.36 & 0.12 & 0.21 & 0.14 \\
& Bernoulli & 0.006 & 0.03 & 0.49 & 0.16 & 0.22 & 0.09 & 0.09 & 0.03  \\
& Normal & 0.005 & 0.01 & 0.08 & 0.37 & 0.04 & 0.04 & 0.12 & 0.05  \\
& Neg Bin & 0.017 & 0.06 & 1.98 & 0.46 & 4.32 & 0.57 & 1.33 & 1.01  \\
\cmidrule(lr){2-10}
& Average & 0.008 & 0.03 & 0.74 & 0.33 & 1.24 & 0.20 & 0.44 & 0.31\\
\midrule
\multirow{5}{*}{n=2000} & Gamma & 0.02 & 0.23 & 2.56 & 4.20 & 3.48 & 5.32 & 1.00 & 0.79 \\
& Bernoulli & 0.01 & 0.33 & 2.31 & 1.03 & 1.48 & 5.46 & 0.22 & 0.09 \\
& Normal & 0.01 & 0.14 & 0.42 & 3.60 & 0.09 & 5.16 & 0.32 & 0.19\\
& Neg Bin & 0.07 & 0.45 & 16.57 & 7.12 & 30.95 & 10.04 & 6.71 & 6.12  \\
\cmidrule(lr){2-10}
& Average & 0.03 & 0.29 & 5.47 & 3.99 & 9.00 & 6.49 & 2.06 & 1.80 \\
\midrule
\multirow{5}{*}{\shortstack{Ratio of \\ n=2000 \\/ n=200}} & Gamma & 3.2 & 6.3 & 6.1 & 12.4 & 9.7 & 46.2 & 4.8 & 5.8 \\
& Bernoulli & 1.8 & 10.9 & 4.8 & 6.4 & 6.7 & 57.6 & 2.6 & 3.4 \\
& Normal & 2.0 & 11.2 & 5.0 & 9.8 & 2.1 & 132.8 & 2.8 & 3.5  \\
& Neg Bin & 4.2 & 8.0 & 8.4 & 15.6 & 7.2 & 17.8 & 5.0 & 6.1  \\
\cmidrule(lr){2-10}
& Average & 3.6 & 8.6 & 7.4 & 12.0 & 7.3 & 31.9 & 4.7 & 5.9  \\
\bottomrule
\end{tabular}

\caption{Seconds runtime comparison for each of the fitted models for 100 simulations of two different parameter combinations of each simulated distribution. Runtime is presented for sample size of n=200 and n=2000. A ratio of the total runtime between the sample sizes is included to provide an indication of model runtime scaling with sample size. Linear scaling in runtime would be a 10x ratio.}
\label{tab:runtime_comparison}
\end{table}

Although these total run-time values are not particularly impactful for an individual model, while building and comparing multiple models, and as sample sizes increase in the era of big data, these computational differences become increasingly more noticeable. 

These operations were performed on Fedora Linux 41 with an AMD Ryzen 7 7840U with 64.0 GB of RAM. The following R packages were used for estimation: \pkg{glm} \citep{rstats_glm}, \pkg{glmtoolbox}\citep{vanegas_generalized_2023}, \pkg{gamlss} \citep{gamlsspack}, \pkg{lme4} \citep{lme4pack}, \pkg{gamlss.mx}\citep{gamlss.mx}, \pkg{mgcv}\citep{wood2001mgcv} and \pkg{GJRM} \citep{GJRM2023} in \textbf{R} version 4.4.2.

\section{Application}

We introduce a dataset with distributional properties similar to those described in the simulations. In this setting, compared to the simulation setting, the marginal distributions are not known in advance, so their parameters must be estimated and fit to be able to analyze the dependence structure. The shape of the dependence structure is then interrogated, and a copula of best fit is selected. 

\subsection{Doctor visits in RAND Health and Retirement Longitudinal Study}

The data set considered is from the RAND Health and Retirement Study\citep{rand1}\citep{rand2}, which is a large longitudinal study of different aspects related to population aging in the United States. The HRS (Health and Retirement Study) is sponsored by the National Institute on Aging (grant number NIA U01AG009740) and is conducted by the University of Michigan. An outcome measure of interest is the number of visits to a doctor during the two years between surveys. We choose the two most recent survey time points as an example case, 2018 (time 1) and 2020/21 (time 2). 

\begin{figure}
    \centering
    \includegraphics[width=\linewidth]{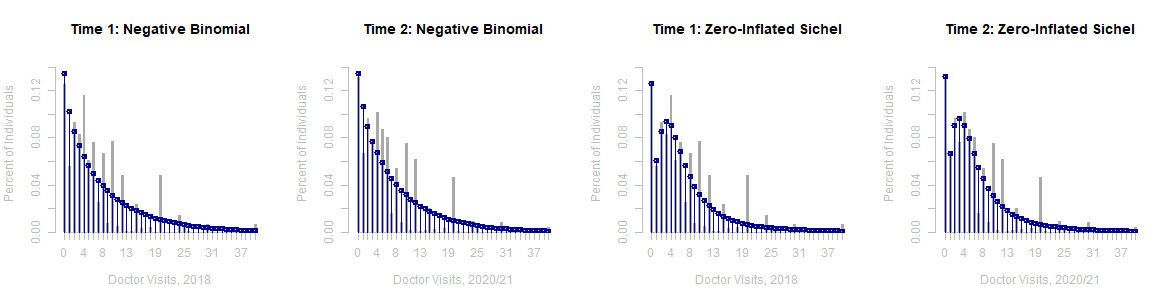}
    \caption{RAND Health and Aging study: Marginal distributions for time 1 and time 2 with fitted negative binomial distributions and fitted Zero Inflated Sichel distributions.}
    \label{fig:applications_rand_margins}
\end{figure}

The sample used consists of 2,404 individuals with observations at both time points. The mean number of doctor visits at time 1 is 8.6 and at time 2 is 7.7. The distribution of number of visits is highly skewed with marginal skewness of 13.7 and 5.7 at times 1 and 2 respectively. Figure \ref{fig:applications_rand_margins} shows the marginal distributions, which appear zero inflated with modes at zero and around five. The best-fit marginal distributions were interrogated using GAMLSS software and it was found that, using the AIC criterion, the zero-inflated Sichel (ZIS) distribution\cite{rigby2019distributions} was the best fit for both marginal distributions. Marginal negative binomial and zero-inflated Sichel distribution fits are shown fitted against the two margins. It can be seen that the zero inflation is not accurately captured by the standard negative binomial. The number of visits is correlated, with a reasonably strong rank correlation measured by Kendall's $\tau$ of 0.40, and weaker Pearson correlation of 0.28. 

We perform a regression for the number of doctor visits. The same models specified in previous sections are fit, with covariates $x_1$ = sex and $x_2$ = age. Results are shown for both a negative binomial distributional fit in Table \ref{tbl:applications_RAND}, which is available to all modelling packages, and ZIS distributional fit in Table \ref{tbl:applications_RAND_flexible}, which is only available in GAMLSS-based methods of which we include GAMLSS (4) and GJRM with the best fitting copula, Frank. 

Table \ref{tbl:applications_RAND} shows the results for the models fit with a \textbf{negative binomial marginal distribution}.
As seen in simulations, it is immediately clear that all the parametric GLMMs (LME4, GAMLSS (4) and GAMM) provide significantly different estimates for the intercepts at times 1 and 2. In simulations, substantially differing estimates from GLMMs were generally biased. Estimates for $\beta_{x_1}$ are all comparable, but for $\beta_{x_2}$, all parametric GLMMs provide approximately twice the effect size of other comparable models, with confidence intervals which would strongly exclude the alternative effect size. These suggest quite different models overall. 

In terms of SE, surprisingly, even though the LME4 and GAMM models provide different parameter estimates, their SEs are quite similar to the other models that adjust for correlation. In contrast, the GAMLSS (4) and GAMLSS NP models provide substantially lower SE than the comparable models, suggesting greater confidence in the estimates. However, in simulations, standard errors for the GLMM for the negative binomial that were lower than the GLM for the coefficients were inconsistent with the true standard error.

In terms of model selection for the negative binomial models, the GAMLSS (4) model provides the best fit by AIC, however, by GAIC (4) and by BIC, the GJRM with the Frank copula provides the best fit. By AIC, the GJRM also provides the second best fit, outperforming LME4. The GEE also appears to provide a similar level of fit to the GJRM based solely on coefficient estimates and standard errors.

\begin{table*}
\centering
\setlength{\tabcolsep}{3.5pt}
\begin{tabular}{l@{\hspace{8pt}}cccc@{\hspace{8pt}}cccc@{\hspace{8pt}}ccccc@{\hspace{8pt}}c}
\toprule
& \multicolumn{4}{c}{Estimates} & \multicolumn{4}{c}{Standard Errors} & \multicolumn{3}{c}{Model Selection Criteria} & \\
\cmidrule(lr){2-5} \cmidrule(lr){6-9} \cmidrule(lr){10-12}
Model & $\hat{\beta_{1}}$ & $\hat{\beta_{2}}$ & $\hat{\beta_{x_1}}$ & $\hat{\beta_{x_2}}$ & $\hat\beta_{1}$ & $\hat\beta_{2}$  & $\hat\beta_{x_1}$ & $\hat\beta_{x_2}$ & AIC & GAIC (4) & BIC & EDF\\
\midrule
GLM       & 1.60 & 1.49 & 0.20 & 0.06 & 0.11 & 0.11 & 0.034 & 0.015 & 30239 & 30247    & 30244   & 5    \\
GEE       & 1.60 & 1.49 & 0.20 & 0.06 & 0.16 & 0.17 & 0.056 & 0.023 & -     & -        & -       & 6    \\
GAMLSS (4)& 0.77 & 0.68 & 0.18 & 0.11 & 0.06 & 0.06 & 0.018 & 0.008 & \textbf{27649} & 31117    & 30044   & 1734 \\
GAMLSS NP & 1.33 & 1.22 & 0.20 & 0.09 & 0.10 & 0.10 & 0.031 & 0.014 & 30031 & 30045    & 30041     & 7    \\
LME4      & 0.70 & 0.61 & 0.20 & 0.13 & 0.14 & 0.14 & 0.044 & 0.020 & 30624 & 32117    & 31655   & 747  \\
GAMM      & 0.59 & 0.50 & 0.20 & 0.14 & 0.14 & 0.14 & 0.045 & 0.020 & -     & -        & -       & -    \\
GJRM (C)  & 1.65 & 1.93 & 0.17 & 0.03 & 0.15 & 0.16 & 0.039 & 0.018 & 29558 & 29576    & \textbf{29570}   & 9    \\
GJRM (N)  & 1.75 & 1.90 & 0.12 & 0.03 & 0.14 & 0.15 & 0.035 & 0.017 & 29587 & \textbf{29605}    & 29600   & 9    \\
GJRM (J)  & 1.59 & 1.73 & 0.18 & 0.05 & 0.15 & 0.15 & 0.036 & 0.017 & 29937 & 29955    & 29950   & 9    \\
GJRM (G)  & 1.69 & 1.86 & 0.17 & 0.04 & 0.15 & 0.15 & 0.037 & 0.017 & 29732 & 29750    & 29744   & 9    \\
GJRM (F)  & 1.62 & 1.89 & 0.16 & 0.04 & 0.15 & 0.15 & 0.041 & 0.019 & \textbf{29398} & \textbf{29416}    & \textbf{29411}   & 9    \\
\bottomrule
\end{tabular}
\caption{RAND Health and Aging study: regression results for negative binomial models estimating number of doctor visits for individuals in the two years prior to survey date. Best two model selection results in each column are bolded for ease of reference. $\beta_{x_2}$ is scaled per 10 units. }
\label{tbl:applications_RAND}
\end{table*}

Table \ref{tbl:applications_RAND_flexible} provides results for fitting the \textbf{zero-inflated Sichel (ZIS) response distribution} available only for the GAMLSS and GJRM models. The ZIS distribution is not currently implemented in the GJRM package, however we have manually estimated results for a model that approximates the GJRM model by fitting two GAMLSS ZIS margins, combined with a best fit copula using the package \pkg{VineCopula}\citep{czado_vine_2022}, and calculated standard errors based on simulation.

Across the board, the more flexible GAMLSS models with ZIS marginal distributions provide much better fit by AIC than the best fit exponential family distribution, the negative binomial, for the same data. The GAMLSS model with ZIS distribution has 1,340 lower AIC than the negative binomial GAMLSS fit, and the GJRM with ZIS margin and Frank copula has 638 lower AIC than the best exponential family GJRM fit. This is unsurprising given the clear zero-inflation present in the dataset. In comparing the two ZIS models, GAMLSS and GJRM, we find that the GAMLSS ZIS model is preferred by AIC, while the GJRM ZIS model is preferred by GAIC (4) and BIC due to the high effective degrees of freedom for the GAMLSS ZIS model.

Coefficient estimates for $x_1$ and $x_2$ are very similar for the two ZIS models, however, importantly, the standard error estimates for the GAMLSS (4) are substantially lower than the GJRM, as was the case in simulations. In simulations, low SE values from the GAMLSS model were generally inconsistent with the other random effect models and were substantially lower than the true values. This poses a concern for interpretation of these values, particularly as they result in tighter confidence intervals. For example, $\hat\beta_{x_1}$ is extremely significant ($p<.00001$) based on the GAMLSS model but only just significant ($p=0.02$) for the GJRM. 

Overall, the best-fitting models by likelihood-based criteria are the GAMLSS (4) model and the GJRM for both the negative binomial and zero-inflated Sichel distributional fits. Between the two models, model selection relies on the degrees of freedom penalty, and preference for interpretation. The GJRM model is selected as the best model by likelihood criteria for penalty of 4 or greater for degrees of freedom, otherwise the GAMLSS (4) model is selected. In addition, the GJRM model can be interpreted directly, while the GAMLSS (4) estimates must be carefully interpreted based on the conditioned random effect term which is transformed by the log link. 

\begin{table*}
\centering
\setlength{\tabcolsep}{3.5pt}
\begin{tabular}{l@{\hspace{8pt}}cccc@{\hspace{8pt}}cccc@{\hspace{8pt}}ccccc@{\hspace{8pt}}c}
\toprule
& \multicolumn{4}{c}{Estimates} & \multicolumn{4}{c}{Standard Errors} & \multicolumn{3}{c}{Model Selection Criteria} & \\
\cmidrule(lr){2-5} \cmidrule(lr){6-9} \cmidrule(lr){10-12}
Model & $\hat{\beta_{1}}$ & $\hat{\beta_{2}}$ & $\hat{\beta_{x_1}}$ & $\hat{\beta_{x_2}}$ & $\hat\beta_{1}$ & $\hat\beta_{2}$  & $\hat\beta_{x_1}$ & $\hat\beta_{x_2}$ & AIC & GAIC (4) & BIC & EDF\\
\midrule
GAMLSS (ZIS) & 0.85 & 0.77 & 0.18 & 0.12 & 0.06 & 0.06 & 0.02 & 0.01 & \textbf{26309} & 29067 & 35241 & 1379 \\
GJRM (ZIS,F) & 1.11 & 1.40 & 0.15 & 0.12 & 0.14 & 0.15 & 0.07 & 0.04 & 28758 & \textbf{28788} & \textbf{28842} & 13   \\

\bottomrule
\end{tabular}
\caption{RAND Health and Aging study: regression results for flexible GAMLSS-based models estimating number of doctor visits for individuals in the two years prior to survey date. Marginal distributions used are Zero-Inflated Sichel (ZIS). *GJRM with ZIS margins and Frank copula is estimated manually, not using the GJRM package directly and instead using GAMLSS marginal fits combined with best fit copula using VineCopula which are separately optimized. Best models by selection result are marked in bold.}
\label{tbl:applications_RAND_flexible}
\end{table*}

\section{Limitations}

The primary limitation of the study is the restriction to comparison of longitudinal datasets in only two dimensions. Ideally, performance of copula methods should be compared for any possible longitudinal model able to be fit by the GLMM or GEE, for example 10, 100 or 1,000 time points. There is some evidence that GLMMs utilizing penalized likelihood may perform poorly in the bivariate case (see Nelder in discussion of Rigby and Stasinopoulos\cite{gamlsspack}), as opposed to higher dimensions, which should be a priority for comparison when approaches for copula longitudinal regression in greater dimensions are developed. In addition, the advantage of the copula approach is more pronounced in more than two dimensions as it can model  a time- or covariate-dependent strength of correlation.

The other limitation of the copula method is that it relies on complete data: every observation must be available for every observational unit. Naturally, this is often not the case for longitudinal studies where individuals may leave a study over time. There is an approach that accounts for censoring in a survival setting that has been developed\citep{dettoni_generalized_2020}, however this is not available more broadly and will be an essential feature for these types of models in more than two dimensions.

Other limitations that we have deliberately excluded to maintain a reasonable scope for the study include:
\begin{itemize}
    \item Sample size: We did not find any evidence of substantial differences between models by sample size in early investigations so did not include this dimension in further review.
    \item Broader range of covariate shapes, e.g. smooth terms: Considering the substantial differences in results between models for more simple factors such as intercept, binary and linear, we did not believe it essential to highlight further covariate fit differences. This could be reviewed in further analysis.
    \item Bayesian methods: There are Bayesian methods for both copula and alternative approaches to longitudinal regression, however, given this comparison already includes seven different models, we restrict our approach to only likelihood and pseudo-likelihood approaches. Future reviews may incorporate Bayesian methods. 
\end{itemize}

\section{Conclusion}
Simulations were run of two-timepoint longitudinal datasets generated from bivariate distributions with normal, Gamma, Bernoulli and negative binomial margins and varying non-standard dependence structures. We compare fits from a copula-based Generalized Joint Regression Model (GJRM) with popular methods for modeling correlated datasets including Generalized Linear Mixed Models (GLMM) implemented in four different packages: \pkg{gamlss}, \pkg{lme4}, \pkg{mgcv}, and \pkg{gamlss.mx}, Generalized Estimating Equations (GEE) implemented in \pkg{glmtoolbox} and Generalized Linear Models (GLMs). We evaluated model performance on the basis of coefficient estimate accuracy, model selection criteria of variogram scores and Bayesian information criteria (BIC), and computational complexity.

We find that all models provide relatively accurate estimates of the coefficients for the normal distribution with identity link, and comparable fit by model selection criteria, excepting the GLM which does not account for correlation.

For non-normal distributions, the Gamma and negative binomial with logarithmic link models, and the Bernoulli with logit link models, we find that the GLMMs with parametric random effects provide highly biased estimates for intercept coefficients with inaccurate standard errors, and slightly biased estimates for covariate coefficients with inaccurate standard errors. In addition, we find that popular GLMM modeling packages \pkg{lme4}, \pkg{gamlss} and \pkg{mgcv} often provide inconsistent coefficient estimates and standard errors for the same model. In general, we find that bias and inconsistencies for the GLMMs is more pronounced the higher the skewness of the marginal distributions and/or the rank correlation between the outcome variables. In contrast to the GLMMs, we find that the GJRM and GEE provide accurate estimates for model parameters and their standard errors across all simulated distributions, as long as the copula distribution is well-fitting (in the GJRM case). 

In terms of overall model fit, we find that the GJRM almost always outperforms the GLMMs and all other models included when compared on the basis of BIC, while providing the second best log-likelihood after the GAMLSS GLMM implementation on average across all distributions. The GJRM provides the best overall fit by weighted variogram score of all included models, closely followed by the GEE. In contrast, the GLMMs provide some of the worst fits by this measure, particularly for the log link models for the Gamma and negative binomial. The fit becomes worse with high levels of rank correlation. The GEE coefficient estimates and standard errors, as well as variogram score performance, suggest they provide a similar level of fit to the GJRM; however they lack the additional tools provided by the likelihood-based structure and the control over the correlation structure provided by the GJRM fit.

We fit all models to a real-world longitudinal study of doctor visits and demonstrate that the GAMLSS (4) and GJRM models provide the best overall fit for an exponential family distribution model. The GEE provides similar coefficient estimates and standard errors. However, when the limitation of the exponential family is removed, the GJRM and GAMLSS models provide vastly superior models compared to the remaining packages that are limited to the exponential family, as they are able to capture the zero-inflation and skew present in the dataset more accurately.

\section{Discussion}

Mixed models are extremely popular as a relatively straightforward method of modeling dependent data, using the mechanism of a univariate framework combined with a stochastic effect that is common across dependent observations. More complex structures, such as hierarchical models, are readily implemented. However, the random effect is an artificial construct and is introduced for the purpose of inducing dependence, indirectly. Generalized joint regression models, on the other hand, use the copula as the correlation-inducing construct. This is an approach which imposes the dependence structure in a more direct and controllable way. Careful modeling of the nature of the dependence structure is made possible with the choice of copula and also the ability to model copula parameters with covariates. (We are not aware of GLMM capability to model the dependence structure with covariates.) GEE models fall somewhere between these two in that they offer a degree of control over the dependence structure; however, we have not demonstrated any advantage of these models over either GLMMs or copulas. The drawback of not having a likelihood-based structure limits the tools available for analysis such as model selection criteria.
 
The results of this study indicate that the application of mixed models with random intercept terms to non-normal correlated data using standard packages without major adjustment provides misleading results for marginal coefficients, and this is exacerbated when marginal distributions are skewed and/or rank correlation is high. In addition, given that our findings show deviation in results between GLMM implementations, we suggest it is essential to compare results from different packages  if the GLMM is the chosen model. More generally, when using a non-identity link function for a random-intercept model, users should be extremely careful in parameter interpretation. Estimates should be thoroughly interrogated, particularly if marginal skewness or rank correlation are high. 

Given the performance of GJRMs in this study both in terms of parameter estimates and fit criteria, they provide a strong potential alternative to GLMMs and GEEs for modeling correlated data with a transparent approach, likelihood-based structure and simpler interpretation than the conditional structure of the GLMM.
The greater flexibility, higher level of transparency, and computational simplicity of generalized joint regression modeling pose significant opportunities for improving the methods used for longitudinal data analysis, once the methods are extended beyond the bivariate case.

\section{Acknowledgements}
We would like to thank the reviewers for their valuable feedback which we believe has substantially improved the quality of the paper and the insights provided. 

\bibliographystyle{unsrtnat}
\bibliography{references}

\begin{thebibliography}{76}
\providecommand{\natexlab}[1]{#1}
\providecommand{\url}[1]{\texttt{#1}}
\expandafter\ifx\csname urlstyle\endcsname\relax
  \providecommand{\doi}[1]{doi: #1}\else
  \providecommand{\doi}{doi: \begingroup \urlstyle{rm}\Url}\fi

\bibitem[Laird and Ware(1982)]{Laird1982}
Nan~M Laird and James~H Ware.
\newblock {Random-Effects Models for Longitudinal Data}.
\newblock \emph{Biometrics}, 38\penalty0 (4):\penalty0 963--974, 1982.

\bibitem[Breslow and Clayton(1993)]{Breslow1993}
N.~E. Breslow and D.~G. Clayton.
\newblock Approximate inference in generalized linear mixed models.
\newblock \emph{Journal of the American Statistical Association}, 88\penalty0 (421):\penalty0 9--25, 1993.

\bibitem[Liang and Zeger(1986)]{Liang1986}
KungYee Liang and Scott~L. Zeger.
\newblock {Longitudinal Data Analysis Using Generalized Linear Models}.
\newblock \emph{Biometrika}, 73\penalty0 (1):\penalty0 13--22, 1986.

\bibitem[Hastie and Tibshirani(1990)]{hastie1990generalized}
Trevor Hastie and Robert Tibshirani.
\newblock \emph{Generalized additive models}.
\newblock CRC Press, Boca Raton, 1990.

\bibitem[Stasinopoulos et~al.(2017)Stasinopoulos, Rigby, Heller, Voudouris, and {De Bastiani}]{Stasinopoulos2017}
D.~M. Stasinopoulos, Robert Rigby, G.Z. Heller, Vlasios Voudouris, and Fernanda {De Bastiani}.
\newblock \emph{{Flexible Regression and Smoothing: Using GAMLSS in R}}.
\newblock CRC Press, Boca Raton, 2017.

\bibitem[Yee and Wild(1996)]{Yee1996}
T~W Yee and C~J Wild.
\newblock {Vector Generalized Additive Models}.
\newblock \emph{Journal of the Royal Statistical Society. Series B (Methodological)}, 58\penalty0 (3):\penalty0 481--493, 1996.

\bibitem[Wild and Yee(1996)]{Wild1996}
C~J Wild and T~W Yee.
\newblock {Additive Extensions to Generalized Estimation Equation Methods}.
\newblock \emph{Journal of the Royal Statistical Society. Series B (Methodological)}, 58\penalty0 (4):\penalty0 711--725, 1996.

\bibitem[McCullagh and Nelder(1989)]{McCullaghNelder89}
P.~McCullagh and J.~A. Nelder.
\newblock \emph{Generalized Linear Models}.
\newblock Chapman \& Hall, London, 2nd edition, 1989.

\bibitem[Vanegas et~al.()Vanegas, Rondón, and Paula]{vanegas_generalized_2023}
L.H. Vanegas, L.M. Rondón, and G.A. Paula.
\newblock Generalized estimating equations using the new r package glmtoolbox.
\newblock 15\penalty0 (2):\penalty0 105--133.

\bibitem[Gory et~al.(2021)Gory, Craigmile, and {MacEachern}]{gory_class_2021}
Jeffrey~J. Gory, Peter~F. Craigmile, and Steven~N. {MacEachern}.
\newblock A class of generalized linear mixed models adjusted for marginal interpretability.
\newblock 40\penalty0 (2):\penalty0 427--440, 2021.

\bibitem[Heagerty and Zeger(2000)]{heagerty_marginalized_2000}
Patrick~J. Heagerty and Scott~L. Zeger.
\newblock Marginalized multilevel models and likelihood inference.
\newblock \emph{Statistical Science}, 15\penalty0 (1):\penalty0 1--19, 2000.
\newblock Publisher: Institute of Mathematical Statistics.

\bibitem[Mills et~al.(2002)Mills, Field, and Dupuis]{mills_marginally_2002}
J.~E. Mills, C.~A. Field, and D.~J. Dupuis.
\newblock Marginally specified generalized linear mixed models: A robust approach.
\newblock \emph{Biometrics}, 58\penalty0 (4):\penalty0 727--734, 2002.

\bibitem[Wang and Louis(2003)]{wang_matching_2003}
Zengri Wang and Thomas~A. Louis.
\newblock Matching conditional and marginal shapes in binary random intercept models using a bridge distribution function.
\newblock \emph{Biometrika}, 90\penalty0 (4):\penalty0 765--775, 2003.
\newblock Publisher: [Oxford University Press, Biometrika Trust].

\bibitem[Molenberghs et~al.(2010)Molenberghs, Verbeke, Demétrio, and Vieira]{molenberghs_family_2010}
Geert Molenberghs, Geert Verbeke, Clarice G.~B. Demétrio, and Afrânio M.~C. Vieira.
\newblock A family of generalized linear models for repeated measures with normal and conjugate random effects.
\newblock \emph{Statistical Science}, 25\penalty0 (3):\penalty0 325--347, 2010.
\newblock Publisher: Institute of Mathematical Statistics.

\bibitem[Kenward and Molenberghs(2016)]{kenward_taxonomy_2016}
Michael~G. Kenward and Geert Molenberghs.
\newblock A taxonomy of mixing and outcome distributions based on conjugacy and bridging.
\newblock \emph{Communications in Statistics - Theory and Methods}, 45\penalty0 (7):\penalty0 1953--1968, 2016.

\bibitem[Molenberghs et~al.(2013)Molenberghs, Kenward, Verbeke, Iddi, and Efendi]{molenberghs_connections_2013}
Geert Molenberghs, Michael Kenward, Geert Verbeke, Samuel Iddi, and Achmad Efendi.
\newblock On the connections between bridge distributions, marginalized multilevel models, and generalized linear mixed models.
\newblock \emph{International Journal of Statistics and Probability}, 2\penalty0 (4):\penalty0 p1, 2013.
\newblock Number: 4.

\bibitem[Sánchez-Miguel et~al.()Sánchez-Miguel, López-Gil, and Tapia-Serrano]{sanchez-miguel_unveiling_2024}
Pedro~Antonio Sánchez-Miguel, José~Francisco López-Gil, and Miguel~Ángel Tapia-Serrano.
\newblock Unveiling the association between 24-hour movement guidelines and academic engagement in adolescents.
\newblock pages 1--6.
\newblock Publisher: Nature Publishing Group.

\bibitem[Anand et~al.()Anand, Koh, Teo, Chin, Mahesh, Chan, Figtree, and Chew]{anand_sex_2024}
Vickram~Vijay Anand, Jaycie Koh, Tobias Teo, Yip~Han Chin, Rishabh Mahesh, Mark~Y. Chan, Gemma~A. Figtree, and Nicholas W.~S. Chew.
\newblock Sex differences in survival following acute coronary syndrome with and without standard modifiable risk factors.

\bibitem[Pi et~al.()Pi, Liu, Jia, Zhang, Liu, Wang, Wang, Li, Ren, and Jin]{pi_periconceptional_2024}
Xin Pi, Chunyi Liu, Xiaoqian Jia, Yali Zhang, Jufen Liu, Bin Wang, Linlin Wang, Zhiwen Li, Aiguo Ren, and Lei Jin.
\newblock Periconceptional polycyclic aromatic hydrocarbon levels in maternal hair and fetal risk for congenital heart defects.
\newblock 286:\penalty0 117251.

\bibitem[Schnall et~al.()Schnall, Scherr, Kuhns, Janulis, Jia, Wood, Almodovar, and Garofalo]{schnall_efficacy_2024}
Rebecca Schnall, Thomas~Foster Scherr, Lisa~M Kuhns, Patrick Janulis, Haomiao Jia, Olivia~R Wood, Michael Almodovar, and Robert Garofalo.
\newblock Efficacy of the {mLab} app: a randomized clinical trial for increasing {HIV} testing uptake using mobile technology.
\newblock page ocae261.

\bibitem[Buch et~al.()Buch, Martin, and Xu]{buch_comfort_2024}
John~R. Buch, Patricia Martin, and Jie Xu.
\newblock Comfort advantages demonstrated with a novel soft contact lens: A randomized clinical trial.
\newblock 10\penalty0 (21).
\newblock Publisher: Elsevier.

\bibitem[Shumba et~al.()Shumba, Fwemba, and Kaymba]{shumba_spatial-temporal_2024}
Samson Shumba, Isaac Fwemba, and Violet Kaymba.
\newblock Spatial-temporal patterns and predictors of timing and inadequate antenatal care utilization in zambia: A generalized linear mixed model ({GLMM}) investigation from 1992 to 2018.
\newblock 4\penalty0 (10):\penalty0 e0003213.
\newblock Publisher: Public Library of Science.

\bibitem[Kerver et~al.()Kerver, Wonderlich, Laam, Amponsah, Nameth, Steffen, Heinberg, Safer, Wonderlich, and Engel]{kerver_naturalistic_2025}
Gail~A. Kerver, Joseph~A. Wonderlich, Leslie~A. Laam, Theresa Amponsah, Katherine Nameth, Kristine~J. Steffen, Leslie~J. Heinberg, Debra~L. Safer, Stephen~A. Wonderlich, and Scott~G. Engel.
\newblock A naturalistic assessment of the relationship between negative affect and loss of control eating over time following metabolic and bariatric surgery.
\newblock 204:\penalty0 107748.

\bibitem[Yu et~al.()Yu, Tang, Fan, Ma, Ye, Cai, Xie, Shi, Baima, Yang, Wang, Jia, and Yang]{yu_associations_2024}
Bin Yu, Wenge Tang, Yunzhe Fan, Chunlan Ma, Tingting Ye, Changwei Cai, Yiming Xie, Yuanyuan Shi, Kangzhuo Baima, Tingting Yang, Yanjiao Wang, Peng Jia, and Shujuan Yang.
\newblock Associations between residential greenness and obesity phenotypes among adults in southwest china.
\newblock 87:\penalty0 103236.

\bibitem[Müllertz et~al.()Müllertz, Stjernqvist, Outzen, Bloch, Elsborg, and Ravn-Haren]{mullertz_cross-sectional_2024}
Alberte Laura~Oest Müllertz, Nanna~Wurr Stjernqvist, Malene~Høj Outzen, Paul Bloch, Peter Elsborg, and Gitte Ravn-Haren.
\newblock A cross-sectional study of the association between food literacy and dietary intake among danish adolescents.
\newblock 200:\penalty0 107526.

\bibitem[Rysavy et~al.()Rysavy, Horbar, Bell, Li, Greenberg, Tyson, Patel, Carlo, Younge, Green, Edwards, Hintz, Walsh, Buzas, Das, Higgins, and {Eunice Kennedy Shriver National Institute of Child Health and Human Development Neonatal Research Network and Vermont Oxford Network}]{rysavy_assessment_2020}
Matthew~A. Rysavy, Jeffrey~D. Horbar, Edward~F. Bell, Lei Li, Lucy~T. Greenberg, Jon~E. Tyson, Ravi~M. Patel, Waldemar~A. Carlo, Noelle~E. Younge, Charles~E. Green, Erika~M. Edwards, Susan~R. Hintz, Michele~C. Walsh, Jeffrey~S. Buzas, Abhik Das, Rosemary~D. Higgins, and {Eunice Kennedy Shriver National Institute of Child Health and Human Development Neonatal Research Network and Vermont Oxford Network}.
\newblock Assessment of an updated neonatal research network extremely preterm birth outcome model in the vermont oxford network.
\newblock 174\penalty0 (5):\penalty0 e196294.

\bibitem[Chopra et~al.()Chopra, Kaatz, Conlon, Paje, Grant, Rogers, Bernstein, Saint, and Flanders]{chopra_michigan_2017}
V.~Chopra, S.~Kaatz, A.~Conlon, D.~Paje, P.~J. Grant, M.~A.~M. Rogers, S.~J. Bernstein, S.~Saint, and S.~A. Flanders.
\newblock The michigan risk score to predict peripherally inserted central catheter‐associated thrombosis.
\newblock 15\penalty0 (10):\penalty0 1951--1962.

\bibitem[Gundelund et~al.()Gundelund, Arlinghaus, Birdsong, Flávio, and Skov]{gundelund_investigating_2022}
Casper Gundelund, Robert Arlinghaus, Max Birdsong, Hugo Flávio, and Christian Skov.
\newblock Investigating angler satisfaction: The relevance of catch, motives and contextual conditions.
\newblock 250:\penalty0 106294.

\bibitem[Casals et~al.()Casals, Girabent-Farrés, and Carrasco]{casals_methodological_2014}
Martí Casals, Montserrat Girabent-Farrés, and Josep~L. Carrasco.
\newblock Methodological quality and reporting of generalized linear mixed models in clinical medicine (2000–2012): A systematic review.
\newblock 9\penalty0 (11):\penalty0 e112653.
\newblock Publisher: Public Library of Science.

\bibitem[Bono et~al.()Bono, Alarcón, and Blanca]{bono_report_2021}
Roser Bono, Rafael Alarcón, and María~J. Blanca.
\newblock Report quality of generalized linear mixed models in psychology: A systematic review.
\newblock 12.
\newblock Publisher: Frontiers.

\bibitem[Bolker et~al.()Bolker, Brooks, Clark, Geange, Poulsen, Stevens, and White]{bolker_generalized_2009}
Benjamin~M. Bolker, Mollie~E. Brooks, Connie~J. Clark, Shane~W. Geange, John~R. Poulsen, M.~Henry~H. Stevens, and Jada-Simone~S. White.
\newblock Generalized linear mixed models: a practical guide for ecology and evolution.
\newblock 24\penalty0 (3):\penalty0 127--135.

\bibitem[Pekár and Brabec()]{pekar_generalized_2018}
Stano Pekár and Marek Brabec.
\newblock Generalized estimating equations: A pragmatic and flexible approach to the marginal {GLM} modelling of correlated data in the behavioural sciences.
\newblock 124\penalty0 (2):\penalty0 86--93.

\bibitem[Thiele and Markussen()]{thiele_potential_2012}
J.~Thiele and B.~Markussen.
\newblock Potential of {GLMM} in modelling invasive spread.
\newblock 2012:\penalty0 1--10.
\newblock Publisher: {CABI}.

\bibitem[Silk et~al.()Silk, Harrison, and Hodgson]{silk_perils_2020}
Matthew~J. Silk, Xavier~A. Harrison, and David~J. Hodgson.
\newblock Perils and pitfalls of mixed-effects regression models in biology.
\newblock 8:\penalty0 e9522.
\newblock Publisher: {PeerJ} Inc.

\bibitem[Arnqvist()]{arnqvist_mixed_2020}
Göran Arnqvist.
\newblock Mixed models offer no freedom from degrees of freedom.
\newblock 35\penalty0 (4):\penalty0 329--335.

\bibitem[Marra and Radice(2017)]{Marra2017}
Giampiero Marra and Rosalba Radice.
\newblock {Bivariate copula additive models for location, scale and shape}.
\newblock \emph{Computational Statistics and Data Analysis}, 112:\penalty0 99--113, 2017.

\bibitem[Nelsen(2007)]{Nelsen2007}
Roger~B Nelsen.
\newblock \emph{{An Introduction to Copulas, Second Edition}}.
\newblock 2007.

\bibitem[Trivedi and Zimmer(2007)]{Trivedi2007}
P~K Trivedi and David~M Zimmer.
\newblock \emph{{Copula modeling : an introduction for practitioners}}.
\newblock Now Publishers, Boston, 2007.

\bibitem[Sklar(1973)]{sklar1973random}
Abe Sklar.
\newblock Random variables, joint distribution functions, and copulas.
\newblock \emph{Kybernetika}, 9\penalty0 (6):\penalty0 449--460, 1973.

\bibitem[Palaro and Hotta(2006)]{ParraPalaro2006}
Helder~Parra Palaro and Luiz~Koodi Hotta.
\newblock {Using Conditional Copula to Estimate Value at Risk}.
\newblock \emph{Journal of Data Science}, 4:\penalty0 93--115, 2006.

\bibitem[Pitt et~al.(2006)Pitt, Chan, and Kohn]{Pitt2006}
Michael Pitt, David Chan, and Robert Kohn.
\newblock {Efficient Bayesian inference for Gaussian copula regression models}.
\newblock \emph{Biometrika}, 93\penalty0 (3):\penalty0 537--554, sep 2006.

\bibitem[Kr{\"{a}}mer et~al.(2013)Kr{\"{a}}mer, Brechmann, Silvestrini, and Czado]{Kramer2013}
Nicole Kr{\"{a}}mer, Eike~C Brechmann, Daniel Silvestrini, and Claudia Czado.
\newblock {Total loss estimation using copula-based regression models}.
\newblock \emph{Insurance: Mathematics and Economics}, 53\penalty0 (3):\penalty0 829--839, 2013.

\bibitem[Patton(2006)]{Patton2006}
Andrew Patton.
\newblock {Modeling asymmetric exchange rate dependence}.
\newblock \emph{International Economic Review}, 47\penalty0 (2):\penalty0 527--556, 2006.

\bibitem[Acar et~al.(2011)Acar, Craiu, and Yao]{Acar2011}
Elif~F Acar, Radu~V Craiu, and Fang Yao.
\newblock {Dependence Calibration in Conditional Copulas: A Nonparametric Approach}.
\newblock \emph{Biometrics}, 67\penalty0 (2):\penalty0 445--453, 2011.

\bibitem[Gijbels et~al.(2011)Gijbels, Veraverbeke, and Omelka]{Gijbels2011}
Ir{\`{e}}ne Gijbels, No{\"{e}}l Veraverbeke, and Marel Omelka.
\newblock {Conditional copulas, association measures and their applications}.
\newblock \emph{Computational Statistics and Data Analysis}, 55\penalty0 (5):\penalty0 1919--1932, 2011.

\bibitem[Acar et~al.(2013)Acar, Craiu, and Yao]{Acar2013}
Elif~F Acar, Radu~V Craiu, and Fang Yao.
\newblock {Statistical testing of covariate effects in conditional copula models}.
\newblock \emph{Electronic Journal of Statistics}, 7:\penalty0 2822--2850, 2013.

\bibitem[Craiu and Sabeti(2012)]{Craiu2012}
Radu~V Craiu and Avideh Sabeti.
\newblock {In mixed company: Bayesian inference for bivariate conditional copula models with discrete and continuous outcomes}.
\newblock \emph{Journal of Multivariate Analysis}, 110:\penalty0 106--120, sep 2012.

\bibitem[Sabeti et~al.(2014)Sabeti, Wei, and Craiu]{Sabeti2014}
Avideh Sabeti, Mian Wei, and Radu~V. Craiu.
\newblock Additive models for conditional copulas.
\newblock \emph{Stat}, 3\penalty0 (1):\penalty0 300--312, 2014.

\bibitem[Marra and Radice(2025)]{marra_copula_2025}
G.~Marra and R.~Radice.
\newblock \emph{Copula Additive Distributional Regression Using R}.
\newblock Chapman \& Hall/{CRC} The R Series. {CRC} Press, 2025.

\bibitem[Vatter and Chavez-Demoulin(2015)]{Vatter2015}
Thibault Vatter and Val{\'{e}}rie Chavez-Demoulin.
\newblock {Generalized additive models for conditional dependence structures}.
\newblock \emph{Journal of Multivariate Analysis}, 141:\penalty0 147--167, 2015.

\bibitem[Stasinopoulos et~al.(2024)Stasinopoulos, Kneib, Klein, Mayr, and Heller]{stasinopoulos_generalized_2024}
M.~D. Stasinopoulos, T.~Kneib, N.~Klein, A.~Mayr, and G.~Z. Heller.
\newblock \emph{Generalized Additive Models for Location, Scale and Shape: A Distributional Regression Approach, with Applications}.
\newblock Cambridge Series in Statistical and Probabilistic Mathematics. Cambridge University Press, 2024.
\newblock ISBN 978-1-00-941006-9.

\bibitem[{R Core Team}(2024)]{R2024}
{R Core Team}.
\newblock \emph{R: A Language and Environment for Statistical Computing}.
\newblock R Foundation for Statistical Computing, Vienna, Austria, 2024.

\bibitem[Klein and Kneib(2016)]{Klein2016}
Nadja Klein and Thomas Kneib.
\newblock {Simultaneous inference in structured additive conditional copula regression models: a unifying Bayesian approach}.
\newblock \emph{Statistics and Computing}, 26:\penalty0 841--860, 2016.

\bibitem[Czado and Nagler(2022)]{czado_vine_2022}
Claudia Czado and Thomas Nagler.
\newblock Vine copula based modeling.
\newblock \emph{Annual Review of Statistics and Its Application}, 9\penalty0 (1):\penalty0 453--477, 2022.

\bibitem[Marra et~al.(2023)Marra, Fasiolo, Radice, and Winkelmann]{Marra2023}
Giampiero Marra, Matteo Fasiolo, Rosalba Radice, and Rainer Winkelmann.
\newblock A flexible copula regression model with {B}ernoulli and {T}weedie margins for estimating the effect of spending on mental health.
\newblock \emph{Health Economics (United Kingdom)}, 32:\penalty0 1305--1322, 6 2023.

\bibitem[Dorn et~al.(2023)Dorn, Radice, Marra, and Kneib]{Dorn2023}
Franziska Dorn, Rosalba Radice, Giampiero Marra, and Thomas Kneib.
\newblock A bivariate relative poverty line for leisure time and income poverty: Detecting intersectional differences using distributional copulas.
\newblock \emph{Review of Income and Wealth}, 0, 2023.

\bibitem[Bates et~al.(2015)Bates, M{\"a}chler, Bolker, and Walker]{lme4pack}
Douglas Bates, Martin M{\"a}chler, Ben Bolker, and Steve Walker.
\newblock Fitting linear mixed-effects models using {lme4}.
\newblock \emph{Journal of Statistical Software}, 67\penalty0 (1):\penalty0 1--48, 2015.

\bibitem[Rigby and Stasinopoulos(2005)]{gamlsspack}
R.~A. Rigby and D.~M. Stasinopoulos.
\newblock Generalized additive models for location, scale and shape.
\newblock \emph{Journal of the Royal Statistical Society Series C: Applied Statistics}, 54:\penalty0 507--554, 2005.

\bibitem[Wood(2017)]{mgcv}
S.N Wood.
\newblock \emph{Generalized Additive Models: An Introduction with R}.
\newblock Chapman and Hall/CRC, 2 edition, 2017.

\bibitem[Kock and Klein(2025)]{Kock2023}
Lucas Kock and Nadja Klein.
\newblock Truly multivariate structured additive distributional regression.
\newblock \emph{Journal of Computational and Graphical Statistics}, 0\penalty0 (0):\penalty0 1--13, 2025.

\bibitem[Genz and Bretz(2009)]{mvtnorm}
Alan Genz and Frank Bretz.
\newblock \emph{Computation of Multivariate Normal and t Probabilities}.
\newblock Lecture Notes in Statistics. Springer-Verlag, Heidelberg, 2009.
\newblock ISBN 978-3-642-01688-2.

\bibitem[Stein and Juritz()]{stein_bivariate_1987}
Gillian~Z. Stein and June~M. Juritz.
\newblock Bivariate compound {P}oisson distributions.
\newblock \emph{Communications in Statistics - Theory and Methods}, 16\penalty0 (12):\penalty0 3591--3607.

\bibitem[Rigby et~al.(2019)Rigby, Stasinopoulos, Heller, and De~Bastiani]{rigby2019distributions}
R.~A. Rigby, M.~D. Stasinopoulos, G.~Z. Heller, and F.~De~Bastiani.
\newblock \emph{Distributions for Modeling Location, Scale, and Shape: Using GAMLSS in R}.
\newblock Chapman \& Hall/CRC, Boca Raton, 2019.

\bibitem[Nadarajah and Gupta(2006)]{Nadarajah2006}
Saralees Nadarajah and Arjun~K. Gupta.
\newblock {Some bivariate gamma distributions}.
\newblock \emph{Applied Mathematics Letters}, 19\penalty0 (8):\penalty0 767--774, 2006.

\bibitem[Marra and Radice(2023)]{GJRM2023}
Giampiero Marra and Rosalba Radice.
\newblock \emph{GJRM: Generalised Joint Regression Modelling}, 2023.
\newblock R package version 0.2-6.2.

\bibitem[{R Core Team}(2023)]{rstats_glm}
{R Core Team}.
\newblock \emph{R: A Language and Environment for Statistical Computing}.
\newblock R Foundation for Statistical Computing, Vienna, Austria, 2023.

\bibitem[Wood(2001)]{wood2001mgcv}
Simon~N Wood.
\newblock mgcv: Gams and generalized ridge regression for r.
\newblock \emph{R news}, 1\penalty0 (2):\penalty0 20--25, 2001.

\bibitem[Stasinopoulos and Rigby(2024)]{gamlss.mx}
Mikis Stasinopoulos and Bob Rigby.
\newblock \emph{gamlss.mx: Fitting Mixture Distributions with GAMLSS}, 2024.
\newblock R package version 6.0-1.

\bibitem[Donohue et~al.(2011)Donohue, Overholser, Xu, and Vaida]{donohue_conditional_2011}
M.~C. Donohue, R.~Overholser, R.~Xu, and F.~Vaida.
\newblock Conditional akaike information under generalized linear and proportional hazards mixed models.
\newblock \emph{Biometrika}, 98\penalty0 (3):\penalty0 685--700, 2011.

\bibitem[Gneiting and Raftery()]{gneiting_strictly_2007}
Tilmann Gneiting and Adrian~E Raftery.
\newblock Strictly proper scoring rules, prediction, and estimation.
\newblock 102\penalty0 (477):\penalty0 359--378.

\bibitem[noa()]{noauthor_variogram-based_nodate}
Variogram-based proper scoring rules for probabilistic forecasts of multivariate quantities in: Monthly weather review volume 143 issue 4 (2015).

\bibitem[ran(2023{\natexlab{a}})]{rand1}
{RAND HRS Longitudinal file 2020 (V1). Produced by the RAND Center for the Study of Aging, with funding from the National Institute on Aging and the Social Security Administration}, 2023{\natexlab{a}}.

\bibitem[ran(2023{\natexlab{b}})]{rand2}
{Health and Retirement Study, (RAND HRS Longitudinal file 2020 (V2)) public use dataset. Produced and distributed by the University of Michigan with funding from the National Institute on Aging (grant number NIA U01AG009740)}, 2023{\natexlab{b}}.

\bibitem[Dettoni et~al.()Dettoni, Marra, and Radice]{dettoni_generalized_2020}
Robinson Dettoni, Giampiero Marra, and Rosalba Radice.
\newblock Generalized link-based additive survival models with informative censoring.
\newblock 29\penalty0 (3):\penalty0 503--512.
\newblock ISSN 1061-8600.

\bibitem[Yeo and Milne(1991)]{Yeo1991}
G.~F. Yeo and R.~K. Milne.
\newblock {On characterizations of beta and gamma distributions}.
\newblock \emph{Statistics and Probability Letters}, 11\penalty0 (3):\penalty0 239--242, mar 1991.

\bibitem[Abramowitz and Stegun(1972)]{Abramowitz1972}
Milton Abramowitz and Irene~A Stegun.
\newblock \emph{{Handbook of Mathematical Functions}}.
\newblock Washington, D.C., 1972.

\end{thebibliography}

\appendix

\section{Bivariate gamma distribution used for simulation studies} \label{sec:appendix biv gamma}

The bivariate Gamma used\citep{Nadarajah2006} induces dependence between two Gamma distributions by multiplying by a common Beta-distributed andom variable, with some restriction on the choice of parameters. The approach is as follows:
\begin{enumerate}
\item Generate three independent random variables:
\begin{align*}
W &\sim \text{Beta}(\alpha,\beta)\\
U &\sim \text{Gamma}(\alpha+\beta,1/\mu'_1)\\
V &\sim \text{Gamma}(\alpha+\beta,1/\mu'_2)
\end{align*}
\item Let $Y_1=WU$ and $Y_2=WV$. Multiplication by $W$ induces dependence between $Y_1$ and $Y_2$.
\end{enumerate}

The properties of the bivariate Gamma distribution of $Y_1$ and $Y_2$, with parameters $\mu'_1>0, \mu'_2>0, \alpha>0$ and $\beta>0$\citep{Yeo1991} are as follows. The marginal distributions are:
$$Y_1\sim \text{Gamma}(\alpha,1/\mu'_1),\ \E(Y_1)=\alpha/\mu'_1$$
$$Y_2\sim \text{Gamma}(\alpha,1/\mu'_2),\ \E(Y_2)=\alpha/\mu'_2$$
$$$$
with correlation coefficient 
$$\text{corr}(Y_1,Y_2) = \rho = \frac{\sqrt{\alpha\beta}}{\alpha +\beta+1}.$$
The parameters $\alpha$ and $\beta$ fully define the dependence structure between the two marginal Gamma distributions. The bivariate distribution is described by its probability density function (PDF):
\begin{gather}
f(y_1,y_2)=C\Gamma(\beta)(y_1y_2)^{\alpha +\beta-1}
	\left(\frac{y_1}{\mu_1}+\frac{y_2}{\mu_2}\right)^{\frac{\alpha-1}{2}-(\alpha +\beta)} 
 \exp\left[-\frac{1}{2}\left(\frac{y_1}{\mu_1}+\frac{y_2}{\mu_2}\right)\right]
 W_{\alpha+\frac{1-\alpha}{2},\alpha +\beta-\frac{\alpha}{2}}\left(\frac{y_1}{\mu_1}+\frac{y_2}{\mu_2}\right), \nonumber
 \label{bivgammapdf}
\end{gather}

for $y_1>0$,$y_2>0$ and where the constant $C$ is given by
$$\frac{1}{C}=(\mu_1\mu_2)^{\alpha +\beta}\Gamma(\alpha +\beta)\Gamma(\alpha)\Gamma(\beta)$$

and $W_{\lambda,\mu}$ is the Whittaker function \citep{Abramowitz1972}:
\begin{gather*}
 W_{\lambda,\mu}(p)= \frac{p^{\mu+\frac{1}{2}} \exp(-p/2) }{\Gamma(\mu-\lambda+1/2)} 
	\int_{0}^{\infty} t^{\mu-\lambda-1/2} (1+t)^{\mu+\lambda-1/2}\exp(-pt)dt.
\end{gather*}
All the dependence structures from this bivariate Gamma incorporate a skew towards lower value dependence with differing strength of overall dependence, and a differing extent to which higher value dependence is also exhibited. The highest rank correlation within the dependence structure is achieved when $\alpha$ is low and $\beta$ is high.

The second parameter of the multiplicative beta distribution, $\beta$, has no effect on the marginal distributions of $Y_1$ and $Y_2$ but is an important component in defining the dependence between the two variables. $\alpha$ and $\beta$ together determine the strength of the rank correlation of the dependence structure. $\alpha$ has the added effect of increasing marginal skewness with lower values. 

We reparameterize this distribution to align with the GAMLSS parametrization of the Gamma distribution\citep{rigby2019distributions}. We define $\mu_t$ to capture marginal means, that is, $\mu_t=\alpha/\mu'_t$, and define $\sigma=1/\sqrt{\alpha}$. We rename $\beta$ to avoid confusion with the parameter used elsewhere in the text, and align with the notation used for the copula parameters, $\theta=\beta$. The bivariate Gamma in our notation is then
\begin{align*}
  \begin{pmatrix}    
Y_1\\Y_2
\end{pmatrix}\sim \text{BivGamma}(\mu_1, \mu_2, \sigma, \theta)  
\end{align*}
where the marginal distributions are Gamma distributions in the GAMLSS (\texttt{GA}) parametrization:
\begin{align*}
Y_t&\sim\texttt{GA}(\mu_t,\sigma),\quad \text{E}(Y_t)=\mu_t, \quad\text{Var}(Y_t)=\mu_t^2\sigma^2\qquad\qquad\text{for } t=1,2\\
\text{and}\qquad\qquad\qquad\qquad\qquad\qquad& \\
\text{corr}(Y_1, Y_2)&=\frac{\sigma\sqrt{\theta}}{1+\sigma^2+\sigma^2\theta}.
\end{align*}

\section{Maximum likelihood estimators for exponential family log link model parameters} \label{sec:appendix_MLE}

For the exponential family we write the log-likelihood as:
$$\log\left[f(y_i|\mu,\phi)   \right] = \frac{y_i \log(\mu)-\mu}{\phi}+\log\left[ c(y_i,\phi) \right]$$
We investigate the maximum likelihood estimators for the GLM under independence, and GLMM conditional on random effects, and under two alternative parametrizations. 

\textbf{GLM, independence, original parametrization:} For the \textbf{marginal model} with parameters $\mu_1$ and $\mu_2$, where $\log(\mu_1)=\beta_1$ and $\log(\mu_2)=\beta_2$, we find the maximum likelihood estimators as follows:
\begin{align*}
    \mathcal{L} = \sum_{i=1}^n \log\left[f(y_i|\mu_1,\mu_2,\phi)   \right]&=\sum_{i=1}^n \frac{y_{i1}\beta_1-e^{\beta_1}+y_{i2}\beta_2-e^{\beta_2}}{\phi}+\log\left[ c(y_i,\phi) \right]\\
    \frac{\partial \mathcal{L}}{\partial\beta_1}&=\sum_{i=1}^n\frac{y_{i1}-e^{\beta_1}}{\phi}=0 \implies \hat{\beta_1}=\log(\overline{y}_1 ) \\
    \frac{\partial \mathcal{L}}{\partial\beta_2}&=\sum_{i=1}^n\frac{y_{i2}-e^{\beta_2}}{\phi}=0 \implies \hat{\beta_2}=\log(\overline{y}_2)
\end{align*}

\textbf{GLMM, original parametrization:} For the \textbf{conditional model} with parameters $\mu_1$ and $\mu_2$, where $\log(\mu_{i1})=\beta_1+b_i$ and $\log(\mu_{i2})=\beta_2+b_i$, we find the maximum likelihood estimators as follows:
\begin{align*}
    \mathcal{L} = \sum_{i=1}^n \log\left[f(y_i|\mu_1,\mu_2,\phi, b_i)   \right]&=\sum_{i=1}^n \frac{y_{i1}(\beta_1+b_i)-e^{\beta_1+b_i}+y_{i2}(\beta_2+b_i)-e^{\beta_2+b_i}}{\phi}+\log\left[ c(y_i,\phi) \right]\\
    \frac{\partial \mathcal{L}}{\partial\beta_1}&=\sum_{i=1}^n\frac{y_{i1}-e^{\beta_1+b_i}}{\phi}=0 \implies \hat{\beta_1}=\log(\overline{y}_1 )-\log\left( \sum_{i=1}^n b_i \right) \\
    \frac{\partial \mathcal{L}}{\partial\beta_2}&=\sum_{i=1}^n\frac{y_{i2}-e^{\beta_2+b_i}}{\phi}=0 \implies \hat{\beta_2}=\log(\overline{y}_2 )-\log\left( \sum_{i=1}^n b_i \right) 
\end{align*}

\textbf{GLM, independence, reparametrized:} For the \textbf{marginal model }with parameters $\beta_1$ and $\beta_t$ where $\log(\mu_1)=\beta_1$ and $\log(\mu_2)=\beta_1+\beta_t$:
\begin{align*}
        \mathcal{L} = \sum_{i=1}^n \log\left[f(y_i|\beta_1,\beta_t,\phi)   \right]&=\sum_{i=1}^n \frac{y_{i1}(\beta_1)-e^{\beta_1}+y_{i2}(\beta_1+\beta_t)-e^{\beta_1+\beta_t}}{\phi}+\log\left[ c(y_i,\phi) \right]\\
        \frac{\partial \mathcal{L}}{\partial\beta_t}&=\sum_{i=1}^n\frac{y_{i2}-e^{\beta_1+\beta_t}}{\phi}=0 \implies \hat{\beta_1}+\hat{\beta_t}=\log(\overline{y}_2 ) \\
        &\text{substitute in value for } \hat{\beta_1}=\log(\overline{y}_1 ) \\
        \hat{\beta_t}&=\log(\overline{y}_{2})-\log(\overline{y}_{1})
\end{align*}

\textbf{GLMM, reparametrized:} For the \textbf{conditional model }with parameters $\beta_1$ and $\beta_t$ where $\log(\mu_1)=\beta_1+b_i$ and $\log(\mu_2)=\beta_1+\beta_t+b_i$:
\begin{align*}
        \mathcal{L} = \sum_{i=1}^n \log\left[f(y_i|\beta_1,\beta_t,b_i,\phi, b_i)   \right]&=\sum_{i=1}^n \frac{y_{i1}(\beta_1+b_i)-e^{\beta_1+b_i}+y_{i2}(\beta_1+\beta_t+b_i)-e^{\beta_1+\beta_t+b_i}}{\phi}+\log\left[ c(y_i,\phi) \right]\\
        \frac{\partial \mathcal{L}}{\partial\beta_t}&=\sum_{i=1}^n\frac{y_{i2}-e^{\beta_1+\beta_t+b_i}}{\phi}=0 \implies \hat{\beta_1}+\hat{\beta_t}=\log(\overline{y}_2 )-\log\left( \sum_{i=1}^n b_i \right) \\
        &\text{substitute in value for } \hat{\beta_1}=\log(\overline{y}_1 )-\log\left(\sum_{i=1}^n b_i\right) \\
        \hat{\beta_t}&=\log(\overline{y}_{2})-\log(\overline{y}_{1})
\end{align*}

The above shows that the estimator for the conditional model for $\beta_t$ does not depends on $b_i$ and is equivalent to the estimator from the marginal model.

\section{Additional model selection comparison for models} \label{sec:appendix_aic}

\begin{figure*}[h]
\includegraphics[width=\linewidth]{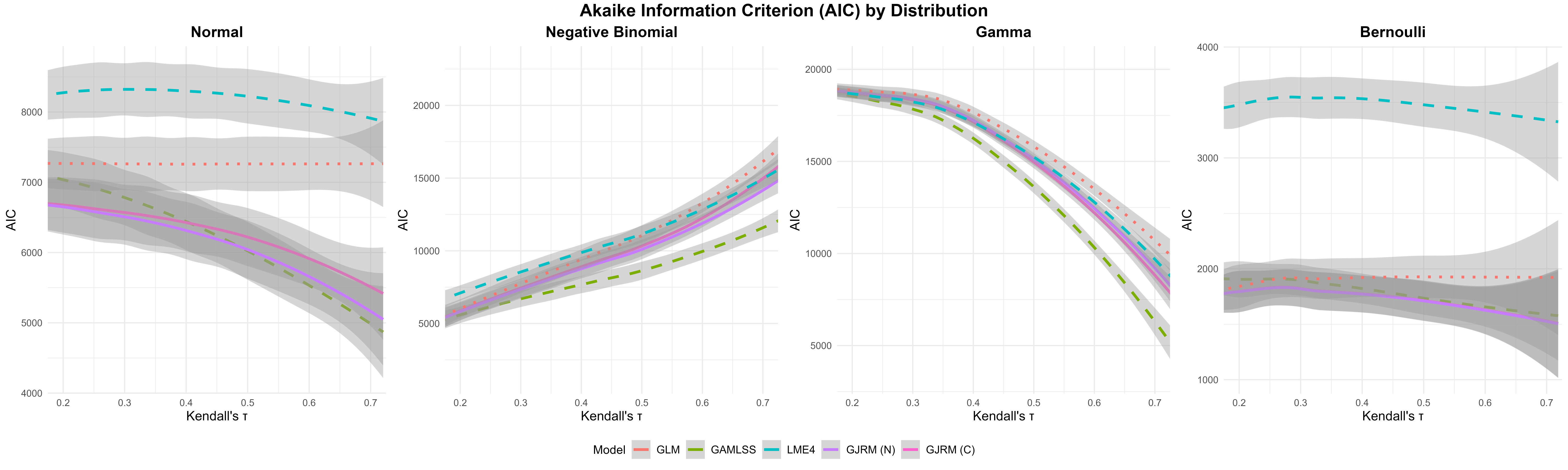} 
 \caption{Average model selection criteria (AIC) for the likelihood-based models across the range of realizations of the simulated distributions. Estimates are plotted against Kendall's $\tau$ which has been computed for each simulation.}
 \label{simLogLik_AIC} 
\end{figure*}

\begin{figure*}[h]
\includegraphics[width=\linewidth]{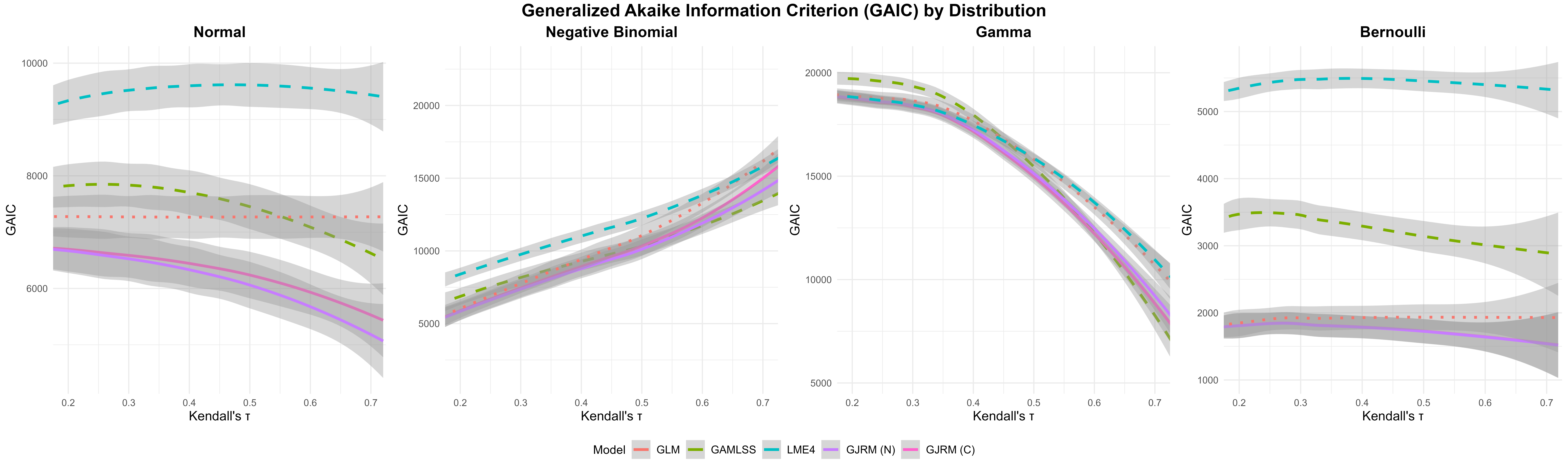} 
 \caption{Average model selection criteria (GAIC (4)) for the likelihood-based models across the range of realizations of the simulated distributions. Estimates are plotted against Kendall's $\tau$ which has been computed for each simulation.}
 \label{simLogLik_AIC} 
\end{figure*}

\begin{figure*} [h]
\includegraphics[width=\linewidth]{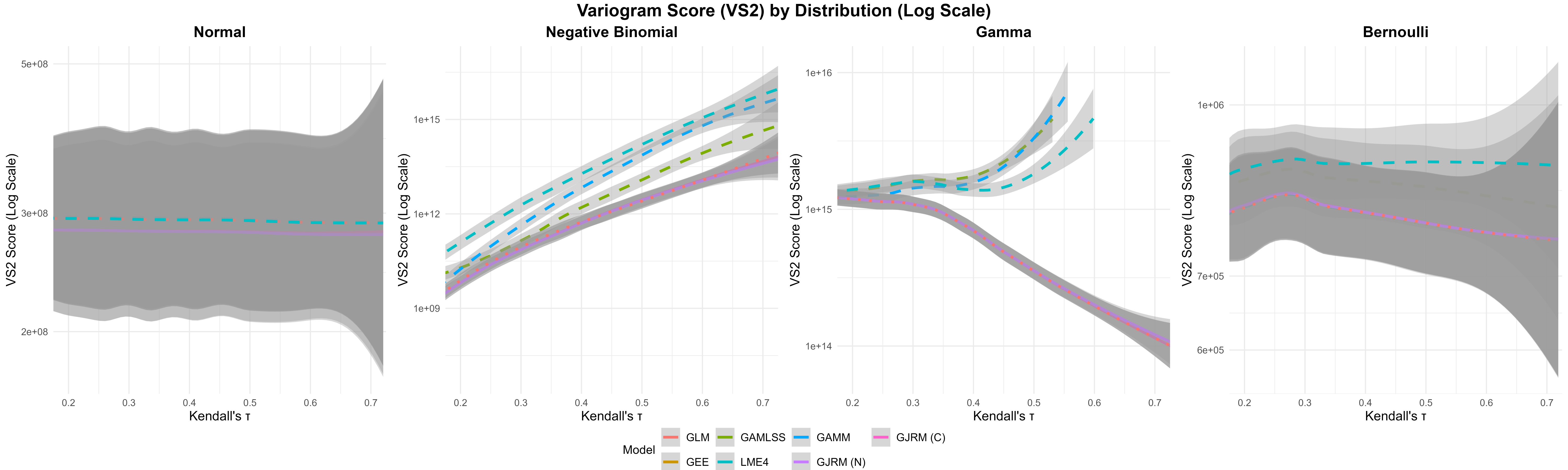} 
 \caption{Average model selection criteria (log of unweighted variogram score) for the seven simulateable models (excluding non-parameteric random effect GAMLSS NP) across the range of realizations of the simulated distributions. Estimates are plotted against Kendall's $\tau$ which has been computed for each simulation.}
 \label{fig:vs2_unweighted} 
\end{figure*}

\section{Coefficients additional results} \label{sec:appendix_coefficients}

\begin{figure*}
\includegraphics[width=\linewidth]{Charts/Mu2_Bias_and_SE_vs_Correlation_Combined_2025-10-23.png}
 \caption{Bias for the estimate of the intercept for time 2 plotted with inverse link transform $g^{-1}({\beta_2})$ (left) and standard error for $\beta_{2}$ (right), across the range of realizations of the simulated distributions, plotted against Kendall's $\tau$. Random effect models (GLMMs) are shown as dashed lines, GJRMs are solid and GLM, GEE are dotted lines for ease of reference.}
 \label{fig:simBiasMu2} 
\end{figure*}

\begin{figure*}
\includegraphics[width=\linewidth]{Charts/X2_Bias_and_SE_vs_Correlation_Combined_2025-10-21.png}
 \caption{Bias for the estimate of the coefficient for $x_2$ (left) and its standard error (right) across the range of realizations of the simulated distributions, plotted against Kendall's $\tau$. Random effect models (GLMMs) are shown as dashed lines, GJRMs are solid and GLM, GEE are dotted lines for ease of reference. Note the GAMLSS model results for the logistic are extremely variable with many extreme values and so cannot be shown entirely on the plot}
 \label{fig:simBiasX2} 
\end{figure*}

\end{document}